\def\isarxiv{1} 

\ifdefined\isarxiv
\documentclass[11pt]{article}
\else
\documentclass[nohyperref]{article}
\usepackage[final]{neurips_2022}
\usepackage{xcolor}
\fi

\usepackage{amsmath}
\usepackage{amsthm}
\allowdisplaybreaks
\usepackage{amssymb}
\usepackage{algorithm}
\usepackage{subfig}
\usepackage{wrapfig,epsfig}
\usepackage{epstopdf}
\usepackage{url}
\usepackage{graphicx}
\usepackage{color}
\usepackage{epstopdf}
\usepackage{algpseudocode}
\usepackage{scrextend}
\usepackage[T1]{fontenc}
\usepackage{bbm}
\usepackage{comment}
\usepackage{multicol}
\usepackage{dsfont}
\usepackage{mathtools}
\usepackage{enumitem}
\usepackage{thmtools,thm-restate}

\usepackage{tikz}
\usepackage{hyperref}
\definecolor{mydarkblue}{rgb}{0,0.08,0.45}
\hypersetup{colorlinks=true, citecolor=mydarkblue,linkcolor=mydarkblue}

\usetikzlibrary{arrows}
\ifdefined\isarxiv
\usepackage[margin=1in]{geometry}
\else 
\fi

\graphicspath{{./figs/}}

\definecolor{b2}{RGB}{51,153,255}
\definecolor{mygreen}{RGB}{80,180,0}

\newcommand{\Lichen}[1]{\textcolor{red}{[Lichen: #1]}}

\theoremstyle{plain}
\newtheorem{theorem}{Theorem}[section]
\newtheorem{lemma}[theorem]{Lemma}
\newtheorem{proposition}[theorem]{Proposition}

\newtheorem{corollary}[theorem]{Corollary}

\theoremstyle{definition}
\newtheorem{definition}[theorem]{Definition}

\newtheorem{remark}[theorem]{Remark}

\newcommand{\wh}{\widehat}
\newcommand{\wt}{\widetilde}
\newcommand{\ov}{\overline}

\renewcommand{\epsilon}{\varepsilon}
\renewcommand{\phi}{\varphi}

\newcommand{\R}{\mathbb{R}}

\renewcommand{\hat}{\wh}
\renewcommand{\bar}{\ov}

\newcommand{\poly}{\mathrm{poly}}
\newcommand{\tr}{\mathrm{tr}}
\newcommand{\Tmat}{{\cal T}_{\mathrm{mat}}}
\newcommand{\nnz}{\mathrm{nnz}}
\newcommand{\diag}{\mathrm{diag}}

\DeclareMathOperator{\rank}{rank}

\newcommand{\E}{\mathbb{E}}

\DeclareMathOperator{\sd}{\mathtt{sd}}
\DeclareMathOperator{\opt}{OPT}
\newcommand {\mat}      [1] {\left[\begin{array}{#1}}
\newcommand {\rix}          {\end{array}\right]}

\makeatletter
\def\blfootnote{\gdef\@thefnmark{}\@footnotetext}
\makeatother

\begin{document}

\ifdefined\isarxiv

\title{Dynamic Tensor Product Regression\footnote{A preliminary version of this paper appeared in the proceedings of 36th Conference on Neural Information Processing Systems (NeurIPS 2022).}}
\author{Aravind Reddy\thanks{\texttt{aravind.reddy@cs.northwestern.edu}. Northwestern University.} 
\and 
Zhao Song\thanks{\texttt{zsong@adobe.com}. Adobe Research.} 
\and Lichen Zhang\thanks{\texttt{lichenz@mit.edu}. MIT.}
}
\date{}

\else 

\title{Dynamic Tensor Product Regression}
\author{Aravind Reddy\thanks{\texttt{aravind.reddy@cs.northwestern.edu}. Northwestern University.} 
\And 
Zhao Song\thanks{\texttt{zsong@adobe.com}. Adobe Research.} 
\And Lichen Zhang\thanks{\texttt{lichenz@mit.edu}. MIT. (Author names in alphabetical order, equal contribution)}
}

\maketitle

\fi

\ifdefined\isarxiv
\begin{titlepage}
  \maketitle
  \begin{abstract}
      In this work, we initiate the study of \emph{Dynamic Tensor Product Regression}. One has matrices $A_1\in \mathbb{R}^{n_1\times d_1},\ldots,A_q\in \mathbb{R}^{n_q\times d_q}$ and a label vector $b\in \mathbb{R}^{n_1\ldots n_q}$, and the goal is to solve the regression problem with the design matrix $A$ being the tensor product of the matrices $A_1, A_2, \dots, A_q$ i.e. $\min_{x\in \mathbb{R}^{d_1\ldots d_q}}~\|(A_1\otimes \ldots\otimes A_q)x-b\|_2$. At each time step, one matrix $A_i$ receives a sparse change, and the goal is to maintain a sketch of the tensor product $A_1\otimes\ldots \otimes A_q$ so that the regression solution can be updated quickly. Recomputing the solution from scratch for each round is very slow and so it is important to develop algorithms which can quickly update the solution with the new design matrix. Our main result is a dynamic tree data structure where any update to a single matrix can be propagated quickly throughout the tree. We show that our data structure can be used to solve dynamic versions of not only Tensor Product Regression, but also Tensor Product Spline regression (which is a generalization of ridge regression) and for maintaining Low Rank Approximations for the tensor product.
  \end{abstract}
  \thispagestyle{empty}
\end{titlepage}
\newpage
\else
  \begin{abstract}
      In this work, we initiate the study of \emph{Dynamic Tensor Product Regression}. One has matrices $A_1\in \mathbb{R}^{n_1\times d_1},\ldots,A_q\in \mathbb{R}^{n_q\times d_q}$ and a label vector $b\in \mathbb{R}^{n_1\ldots n_q}$, and the goal is to solve the regression problem with the design matrix $A$ being the tensor product of the matrices $A_1, A_2, \dots, A_q$ i.e. $\min_{x\in \mathbb{R}^{d_1\ldots d_q}}~\|(A_1\otimes \ldots\otimes A_q)x-b\|_2$. At each time step, one matrix $A_i$ receives a sparse change, and the goal is to maintain a sketch of the tensor product $A_1\otimes\ldots \otimes A_q$ so that the regression solution can be updated quickly. Recomputing the solution from scratch for each round is very slow and so it is important to develop algorithms which can quickly update the solution with the new design matrix. Our main result is a dynamic tree data structure where any update to a single matrix can be propagated quickly throughout the tree. We show that our data structure can be used to solve dynamic versions of not only Tensor Product Regression, but also Tensor Product Spline regression (which is a generalization of ridge regression) and for maintaining Low Rank Approximations for the tensor product.
  \end{abstract}
\fi

\section{Introduction}
The task of fitting data points to a line is well-known as the least-squares regression problem~\cite{s81}, which has a wide range of applications in signal processing~\cite{rgy78}, convex optimization~\cite{b15}, network routing~\cite{m13_flow,m16}, and training neural networks~\cite{bpsw21,szz21}. In this work, we study a generalized version of least-squares regression where the design matrix $A$ is a tensor product of $q$ smaller matrices $A_1\otimes A_2\otimes \ldots \otimes A_q$.

Tensor products have been extensively studied in mathematics and the physical sciences since their introduction more than a century ago by Whitehead and Russell in their Principia Mathematica~\cite{wr1912}. They have been shown to have a humongous number of applications in several areas of mathematics like applied linear algebra and statistics~\cite{van2000ubiquitous}. In particular, machine learning applications include image processing~\cite{nagy2006kronecker}, multivariate data fitting~\cite{golubVanLoan2013Book}, and natural language processing ~\cite{panahi2021shapeshifter}. Furthermore, regression problems involving tensor products arise in surface fitting and multidimensional density smoothing~\cite{eilers2006multidimensional}, and structured blind deconvolution problems~\cite{oh2005}, among several other applications, as discussed in~\cite{dssw18,djssw19}. 

Solving tensor product regression in $\ell_2$ norm~\cite{djssw19, ffg22} to a $(1\pm\epsilon)$ precision takes time $\wt O(\sum_{i=1}^q\nnz(A_i)+\poly(dq/(\delta\epsilon)))$, where each matrix $A_i\in \R^{n_i\times d_i}$ and $d=d_1d_2\ldots d_q$, and $\nnz(A)$ denotes the number of nonzero entries of $A$. However, these algorithms are inherently \emph{static}, meaning that even if there is a sparse (or low-rank) update to a \textit{single} design matrix $A_i$, the algorithm must re-process the changed factor, at a cost of at least $\nnz(A_i)$, and solve again, whereas our data structure sketches only the change. In modern machine learning applications, such static algorithms are not practical due to the fact that data points are always evolving and recomputing the solution from scratch every time is computationally too expensive. Hence, it is important to develop efficient \emph{dynamic algorithms} that can adaptively update the solution. For example, graphs for real-world network data are modeled as the tensor product of a large number of smaller graphs~\cite{leskovec2010kronecker}. Many important problems on these graphs can be solved by regression of the  adjacency and Laplacian matrices of these graphs~\cite{st04}. Real-world network data is always time-evolving and so it is crucial to develop dynamic algorithms for solving regression problems where the design matrix is a tensor product of a large number of smaller matrices. Hence, we ask the following question:

\begin{center}
    {\it Can we design a dynamic data structure for tensor product regression that can handle updates to the design matrix and quickly updates the solution?}
\end{center}

We provide a positive answer to the above question.

At the heart of our data structure is a binary tree that can maintain a succinct sketch of $A_1\otimes \ldots \otimes A_q$ and supports an update to one of the matrices quickly. Such tree structures have been useful in designing efficient sketching algorithms for tensor products of vectors~\cite{akk+20,swyz21}. However, the goal of these works has been in reducing the sketching dimension from an exponential dependence on $q$ to a polynomial dependence on $q$. Their results can be directly generalized to solving tensor product regression \emph{statically} but not dynamically. In this work, we build upon the tree structure to solve the Dynamic Tensor Product Regression problem and other related dynamic problems like Dynamic Tensor Spline Regression and Dynamic Tensor Low Rank Approximation. Our key observation is that every sketch in the tree is linear and the leaf sketches can be cached, so updating the tree is efficient when a single leaf of the tree gets an update: specifically, when $A_i$ changes by $B$, we only compute the leaf sketch of $B$ and add the resulting changes to the nodes which fall on the path from that leaf to the root.

\paragraph{Technical Contributions.}
\begin{itemize}
    \item We design a dynamic tree data structure \textsc{DynamicTensorTree} that maintains a succinct representation of the tensor product $A_1\otimes \ldots \otimes A_q$ and supports efficient updates to any of the $A_i$'s.
    \item Consequently, we develop a dynamic algorithm for solving Tensor Product Regression, when one of the matrices $A_i$'s is updated and we need to output an estimate of the new solution to the regression problem quickly.
    \item We also show that we can use our \textsc{DynamicTensorTree} data-structure in a fast dynamic algorithm to solve the Tensor Product Spline Regression problem, which is a generalization of the classic Ridge Regression problem.
    \item We also initiate the study of \emph{Dynamic Tensor Low Rank Approximation}, where the goal is to maintain a low rank approximation (LRA) of the tensor product with dynamic updates, and show how we can use our \textsc{DynamicTensorTree} data structure to solve this problem.
\end{itemize}

\paragraph{Roadmap.}
In section~\ref{sec:related_work}, we first discuss some related work. Then, we provide notation, some background for our work, and the main problem formulation in section~\ref{sec:preliminaries}. In section~\ref{sec:technical_overview}, we provide a technical overview of our paper. Following that, we provide our dynamic tree data structure in section~\ref{sec:tree_data_structure}. We then show how it can be used for Dynamic Tensor Product Regression, Dynamic Tensor Spline Regression, and Dynamic Tensor Low Rank Approximation in section~\ref{sec:applications}. We end with our conclusion and discuss some future directions in section~\ref{sec:conclusion}.

\section{Related Work}\label{sec:related_work}

\paragraph{Sketching.}
Sketching techniques have many applications in numerical linear algebra, such as linear regression, low-rank approximation \cite{cw13,nn13,mm13,bw14,rsw16,swz17,hlw17,alszz18,bbb+19,ivww19,swz19_neurips1,swz19_neurips2,mw21,wy22,cstz22}, distributed problems \cite{wz16,bwz16}, reinforcement learning \cite{wzd+20,ssx21}, projected gradient descent \cite{xss21}, training over-parameterized neural networks \cite{syz21,szz21}, tensor decomposition \cite{swz19_soda}, clustering \cite{emz21}, cutting plane method \cite{jlsw20}, generative adversarial networks \cite{xzz18}, recommender systems \cite{rrs+22}, and linear programming \cite{lsz19,jswz21,sy21}.

\paragraph{Dynamic Least-squares Regression.}
A dynamic version of the ordinary least-squares regression problem (without a tensor product design matrix) was recently studied in~\cite{jpw22}. In their model, at each iteration, a new row is prepended to the design matrix $A$ and a new coordinate is prepended to the vector $b$. To efficiently maintain the necessary parts of the design matrix and the solution, they make use of a fast leverage score maintenance data structure by \cite{cmp20} and achieve a running time of $\wt O(\nnz(A^{(T)})+\poly(d,\epsilon^{-1}))$ where $T$ is the number of time steps. 

\paragraph{Tensor/Kronecker Product Problems.}
Many machine learning problems involve tensor/Kronecker product computations such as, regression \cite{hlw17,dssw18,djssw19}, low-rank approximation \cite{swz19_soda}, fast kernel computation \cite{swyz21}, semi-definite programming \cite{jklps20,hjstz21}, and training over-parameterized neural networks \cite{bpsw21,szz21}. Apart from solving problems which directly involve tensor/Kronecker products, another line of research focuses on constructing polynomial kernels efficiently so that the computation of various kernel problems can be improved~\cite{akk+20,swyz21}.

\section{Preliminaries}\label{sec:preliminaries}

\subsection{Notation}\label{subsec:notation}
For any natural number $n$, we use $[n]$ to denote the set $\{1,2,\dots,n\}$. We use $\E[\cdot]$ to denote expectation of a random variable if it exists. We use $\mathbf{1}[\cdot]$ and $\Pr[\cdot]$ to denote the indicator function and probability of an event respectively. For any natural numbers $a,b,c$, we use $\Tmat(a,b,c)$ to denote the time it takes to multiply two matrices of sizes $a \times b$ and $b \times c$. For a vector $x$, we use $\|x\|_2$ to denote its $\ell_2$ norm. For a matrix $A$, we use $\|A\|_F$ to denote its Frobenius norm. For a matrix $A$, we use $A^\top$ to denote the transpose of $A$. Given two matrices $A\in \R^{n_1\times d_1}$ and $B\in \R^{n_2\times d_2}$, we use $A\otimes B$ to denote their tensor product (which is also referred to as Kronecker product), i.e., $(A\otimes B)_{i_1+(i_2-1)\cdot n_1,j_1+(j_2-1)\cdot d_1}=A_{i_1,j_1}\cdot B_{i_2,j_2}$. For example, for $2 \times 2$ matrices $A$ and $B$, \[\begin{bmatrix} a_{11} & a_{12}\\ a_{21} & a_{22} \end{bmatrix} \otimes B = \begin{bmatrix} a_{11} B & a_{12} B\\ a_{21} B & a_{22} B \end{bmatrix} = \begin{bmatrix} a_{11}b_{11} & a_{11}b_{12} & a_{12}b_{11} & a_{12}b_{12}\\ a_{11}b_{21} & a_{11}b_{22} & a_{12}b_{21} & a_{12}b_{22} \\ a_{21}b_{11} & a_{21}b_{12} & a_{22}b_{11} & a_{22}b_{12} \\ a_{21}b_{21} & a_{21}b_{22} & a_{22}b_{21} & a_{22}b_{22} \end{bmatrix}\] 

We use $\bigotimes$ to denote tensor product of more than two matrices, i.e., $\bigotimes_{i=1}^q A_i = A_1 \otimes A_2 \otimes \cdots \otimes A_q$. For non-negative real numbers $a$ and $b$, we use $a= (1\pm \epsilon)b$ to denote $(1-\epsilon) \cdot b \leq a \leq (1+\epsilon) \cdot b$. For a real matrix $A$, we use $\sigma_i(A)$ to denote its $i$-th largest singular value. For any function $f$, we use $\wt{O}(f)$ to denote $O( f \poly ( \log f ) )$.

\subsection{Problem Formulation}\label{subsec:problem_formulation}
In this section, we define our task. We start with the \emph{static version} of tensor product regression.
\begin{definition}[Static version]
In the $\ell_2$ Tensor Product Regression problem, we are given matrices $A_1, A_2, \dots, A_q$ where $A_i \in \R^{n_i \times d_i}$ and $b \in \R^{n_1 n_2 \dots n_q}$. Let us define $n \coloneqq n_1n_2\dots n_q$ and $d \coloneqq d_1d_2\dots d_q$. The goal is to output $x \in \R^{d}$ such that we minimize the objective: \[\|(A_1 \otimes A_2 \otimes\dots \otimes A_q) x - b\|_2\]
\end{definition}

Our dynamic version of $\ell_2$ Tensor Product Regression involves updating one of the matrices $A_i$ with an update matrix $B$. We formally define the problem as follows:
\begin{definition}[Dynamic version]
In the Dynamic $\ell_2$ Tensor Product Regression problem, we are given matrices $A_1,A_2,\ldots,A_q$ where $A_i\in \R^{n_i\times d_i}$, a sequence $\{(i_1,B_1),(i_2,B_2),\ldots,(i_T,B_T)\}$ with $i_t \in [q], B_t \in \R^{n_{i_t}\times d_{i_t}}$ for all $t \in [T]$, and the goal is to design a data structure with the following procedures:
\begin{itemize}
    \item {\sc Initialize}: The data structure initializes $A_1,\ldots, A_q$ and maintains an approximation to $\bigotimes_{i=1}^q A_i$.
    \item {\sc Update}: At time step $t$, given an index $i_t$ and an update matrix $B_t$, the data-structure needs to update $A_{i_t} \leftarrow A_{i_t}+B_t$, and maintain an approximation to $A_1\otimes \ldots \otimes (A_{i_t}+B_t)\otimes \ldots \otimes A_q$.
    \item {\sc Query}: The data structure outputs a vector $\wh x\in \R^{d}$ such that
    \begin{align*}
        \|(\bigotimes_{i=1}^q A_i)\wh x-b\|_2 = (1\pm\epsilon) \min_{x\in \R^{d}}~\|(\bigotimes_{i=1}^q A_i) x-b\|_2.
    \end{align*}
\end{itemize}
The sequence $\{(i_t,B_t)\}_{t=1}^T$ is fixed in advance, independently of the randomness of the data structure.
\end{definition}
Our main theorems are for this setting. For sequences chosen by an adaptive adversary, Theorem~\ref{thm:adaptive_tree} keeps the tree a subspace embedding, Theorem~\ref{thm:adaptive_regression} solves regression up to an additional additive error $\xi$ against an adversary that does not see the random bits (there $T$ bounds the number of queries), and Proposition~\ref{prop:query_refresh} answers queries from freshly drawn sketches. For low rank approximation, Theorem~\ref{thm:adaptive_tree} gives Corollary~\ref{cor:adaptive_low_rank}.

Note that although we don't discuss updates to the vector $b$ in our model, they are easy to implement in our data-structure. In addition to the standard regression problem, we also consider the spline regression (which is a generalization of ridge regression) and low rank approximation problems. We will provide the definitions for these problems in their respective sections.
\section{Technical Overview}\label{sec:technical_overview}
The two main design considerations for our data structure are as follows: 1).\ We want a data structure that stores a \emph{sketch} of the tensor product, this has the advantage of having a smaller storage space and a faster time to form the tensor product. 2).\ The update time to one of the matrices $A_i$ should be fast, ideally depending on the size of the change $B$ rather than on the size of $A_i$. With these two goals in mind, we introduce a dynamic tree data structure that satisfies both of them simultaneously, building on the sketch of \cite{akk+20}. Specifically, we use each leaf of the tree to represent $S_i A_i$, a sketch of the base matrix that only has $m$ rows instead of $n$ rows, where $m$ depends polynomially on the small dimension $d$. Each internal node represents a sketch of the Tensor product of its two children. To speed up computation, we adapt tensor-typed sketching matrices to avoid the need to explicitly form the Tensor product of the two children. The root of the tree will store a sketch for the tensor product $\bigotimes_{i=1}^q A_i$.

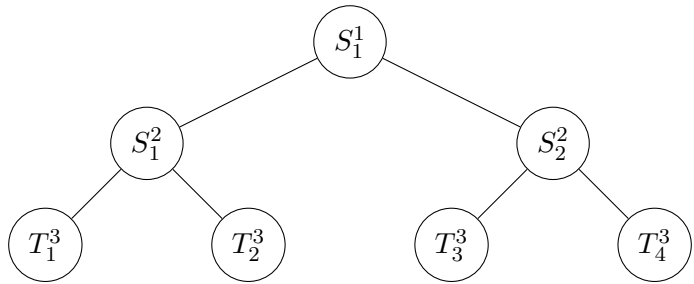
\begin{figure*}[!ht]
    \centering
    \begin{tikzpicture}[scale = 0.9,level distance=1.5cm,
        level 1/.style={sibling distance=6cm},
        level 2/.style={sibling distance=3cm},
        level 3/.style={sibling distance=1.5cm}]
        \node[circle, draw](z) {$S^1_1$}
	        child{node[circle, draw]{$S^2_1$}
		        child{node[circle, draw]{$T^3_1$}}
		        child{node[circle, draw]{$T^3_2$}}
	        }
	        child{node[circle, draw]{$S^2_2$}
		        child{node[circle, draw]{$T^3_3$}}
		        child{node[circle, draw]{$T^3_4$}}
	        };
    \end{tikzpicture}
\caption{A tree for sketching the tensor product of 4 matrices $A_1, A_2, A_3$ and $A_4$. The leaves use sketches of type $T_{\mathrm{base}}$ and internal nodes use sketches of type $S_{\mathrm{base}}$. The algorithm starts to apply $T_i$ to each matrix $A_i$, and then propagates them up the tree.}
    \label{fig:tree}
\end{figure*}

We point out that in order for this tree to work, we need to apply an \emph{independent} sketching matrix to each tree node. Since there are $2q-1$ nodes in total, we will still have a dependence on $q$ in the initialization phase (which will be dominated by the $d$ term). However, during the update phase, our tree has a clear advantage: every sketch in the tree is linear and the leaf sketches are cached, so an update $A_i\gets A_i+B$ only needs $S_iB$ at the leaf, followed by propagating the change along the $O(\log q)$ nodes on the path to the root. The update cost therefore depends on $\nnz(B)$ rather than on $\sum_{i=1}^q \nnz(A_i)$, and this is the key insight for obtaining a faster update time using our dynamic data structure. Moreover, the correctness follows naturally from the guarantee of the tree: as shown in~\cite{akk+20}, the tree itself is an Oblivious Subspace Embedding (OSE) for matrices of proper size, hence, for a sequence of updates fixed in advance, it also preserves the subspace after updating one of the leaves. If the sketches on the updated path are redrawn, the root remains a subspace embedding for the current $\bigotimes_{i=1}^q A_i$ against an adaptive adversary as well, with a larger sketching dimension and an update cost that also depends on $\nnz(A_i)$ (Theorem~\ref{thm:adaptive_tree}). Figure~\ref{fig:tree} is an example tree for tensor product of 4 matrices.

With this data structure, we provide efficient algorithms for dynamic Tensor product regression, dynamic Tensor Spline regression, and dynamic Tensor low rank approximation. For all these problems, we do not explicitly maintain the solutions, rather, we use our tree to quickly update a sketch of the design matrix. In the query phase (when we need to actually solve the regression problems), we use this sketch of updated design matrix to quickly compute the solution.
\section{Dynamic Tree Data Structure}\label{sec:tree_data_structure}
In this section, we introduce our data structure, which contains {\sc Initialize} and {\sc Update} procedures. We'll define task-specific {\sc Query} procedures later. We defer proofs in this section to Appendix~\ref{sec:appendix-tree_data_structure}. We start with an outline of Algorithm~\ref{alg:tree_static}.
\begin{algorithm*}[!t]\caption{Our dynamic tree data structure}\label{alg:tree_static}
\begin{algorithmic}[1]
\State {\bf data structure} {\sc DynamicTensorTree} \Comment{Theorem~\ref{thm:tree_init}}
\State
\State {\bf members}
\State \hspace{4mm} $J_{k,\ell}\in \R^{m \times d_1^{2^\ell}}$ for $0\leq \ell\leq \log q$ and $0\leq k\leq q/2^\ell-1$ 
\Comment{$\ell$ is the level of the tree.}
\State {\bf end members}
\State
\Procedure{Initialize}{$A_1\in \R^{n_1\times d_1},\ldots,A_q\in \R^{n_q\times d_q}, m \in \mathbb{N}_+, C_{\mathrm{base}}, T_{\mathrm{base}}$}
    \For{$\ell=0\to \log q$}
    \For{$k=0\to q/2^\ell-1$}
    \If{$\ell=0$}
    \State Choose a $C_k \in \R^{m \times n_{k+1}}$ from $C_{\mathrm{base}}$. \Comment{$C_{\mathrm{base}}$ can be {\sc CountSketch} or {\sc OSNAP}.}
    \State $J_{k,0}\gets C_k A_{k+1}$
    \Else
    \State Choose a $T_{k,\ell} \in \R^{m \times m^2}$ from $T_{\mathrm{base}}$. \Comment{$T_{\mathrm{base}}$ can be {\sc TensorSketch} or {\sc TensorSRHT}.}
    \State $J_{k,\ell} \gets T_{k,\ell}(J_{2k,\ell-1} \otimes J_{2k+1,\ell-1})$ \Comment{Sketch of tensor product of its children}
    \EndIf
    \EndFor
    \EndFor
\EndProcedure
\State
\Procedure{Update}{$i\in [q], B\in \R^{n_i\times d_i}$}
\For{$\ell=0\to \log q$}
    \State $k_\ell \gets \lfloor (i-1)/2^{\ell} \rfloor$ \Comment{The level-$\ell$ node on the path from leaf $i-1$ to the root}
    \If{$\ell=0$}
    \State $\Delta J_{k_0,0}\gets C_{k_0} B$
    \State $J_{k_0,0}\gets J_{k_0,0} + \Delta J_{k_0,0}$
    \Else
    \If{$k_{\ell-1}$ is even}
    \State $\Delta J_{k_\ell,\ell} \gets T_{k_\ell,\ell} (\Delta J_{2k_\ell,\ell-1} \otimes J_{2k_\ell+1,\ell-1})$ \Comment{The changed child is the left child}
    \Else
    \State $\Delta J_{k_\ell,\ell} \gets T_{k_\ell,\ell} (J_{2k_\ell,\ell-1} \otimes \Delta J_{2k_\ell+1,\ell-1})$ \Comment{The changed child is the right child}
    \EndIf
    \State $J_{k_\ell,\ell} \gets J_{k_\ell,\ell} + \Delta J_{k_\ell,\ell}$
    \EndIf
    \EndFor
\EndProcedure
\end{algorithmic}
\end{algorithm*}


\textbf{Outline of Tree:} For simplicity, let us assume that the number of matrices $q$ is a power of $2$. Also, let us assume that all the matrices are of the same dimension $n_1 \times d_1$. The output is a matrix of dimension $m \times d$ where $m = \poly(q,d,\epsilon^{-2})$. First, we apply $q$ independent base sketch matrices (such as {\sc CountSketch} (Def.~\ref{def:cs}), {\sc OSNAP} (Def.~\ref{def:osnap})) $\{T_i \in \R^{m \times n_1}, i \in [q]\}$ to each of the matrices $A_1, A_2, \dots, A_q$. After this step, all the leaf nodes are of size $m \times d_1$. Then, we have a tree of height $\log q$. Let us use $\ell = 0 \to \log q$ to represent the level of the tree nodes where $\ell = 0$ represents the leaves and $\ell = \log q$ represents the root node. Let us use $J_{k,\ell}\in \R^{m\times d_1^{2^\ell}}$ to denote the matrix stored at the node $k, \ell$. For leaf nodes, we choose base sketches $T_k\in \R^{m\times n_1}$ and compute $J_{k, 0} = T_k A_{k+1}$. At each internal node, we use a sketch for tensor-typed inputs (such as {\sc TensorSketch} (Def.~\ref{def:TensorSketch}) or {\sc TensorSRHT} (Def.~\ref{def:TensorSRHT})), apply it to its two children. The sketching matrix $S_i\in \R^{m\times m^2}$, and it can be applied to its two inputs fast, obtaining a running time of $\wt O(m)$ instead of $O(m^2)$. We inductively compute the tree all the way up to the root for initialization. Note that at the root, we end up with a matrix of size $m\times d$, which is significantly smaller than the tensor product $n\times d$.

\textbf{Leaf sparsity.} Let $s_C$ denote the maximum number of nonzero entries in a column of a leaf sketch $C_k$. Applying $C_k$ to a matrix $X$ costs $O(s_C\nnz(X))$, because each nonzero entry of $X$ meets the $s_C$ nonzero entries of one column of $C_k$. For {\sc CountSketch}, $s_C=1$. For {\sc OSNAP} leaves with {\sc TensorSRHT} internal nodes as in Theorem 2 of~\cite{akk+20}, $s_C=\Theta(\sqrt{q}\cdot{\rm dim}\cdot\log(1/\delta)/\epsilon)$ for the subspace embedding guarantee, where ${\rm dim}$ is the dimension of the subspace to be embedded (see Table~\ref{tab:diff_sketch}), and $s_C=\Theta(\sqrt{q}\log(1/\delta)/\epsilon')$ for approximate matrix product with accuracy $\epsilon'$. For the construction of Theorem 3 of~\cite{akk+20}, $s_C=\wt O(q^4/\epsilon^2)$.

The running time guarantees of Algorithm~\ref{alg:tree_static}, which we will prove in Appendix~\ref{sec:appendix-tree_data_structure}, are as follows:
\begin{theorem}[Running time of our main result. Informal version of Theorem~\ref{thm:app:tree_init}]\label{thm:tree_init}
There is an algorithm (Algorithm~\ref{alg:tree_static}) that has the following procedures:
\begin{itemize}
    \item \textsc{Initialize}($A_1 \in \R^{n_1 \times d_1}, A_2 \in \R^{n_2 \times d_2},\dots, A_q \in \R^{n_q \times d_q}, m\in \mathbb{N}_+, C_{\mathrm{base}}, T_{\mathrm{base}}$): Given matrices $A_1 \in \R^{n_1 \times d_1}, A_2 \in \R^{n_2 \times d_2}, \dots, A_q \in \R^{n_q \times d_q},$, a sketching dimension $m$, families of base sketches $C_{\mathrm{base}}$ and $T_{\mathrm{base}}$, the data-structure pre-processes in time $\wt O(s_C\sum_{i=1}^q \nnz(A_i) + qmd)$, where $s_C$ is the leaf sparsity defined above.
    \item {\sc Update}$(i\in [q], B\in \R^{n_i\times d_i})$: Given a matrix $B$ and an index $i\in [q]$, the data structure updates the approximation to $A_1\otimes \ldots \otimes (A_i+B)\otimes \ldots \otimes A_q$ in time $\wt O(s_C\nnz(B) + md)$.
\end{itemize}
\end{theorem}

\begin{remark}
When every $d_i\geq 2$, we have $d=\prod_{i=1}^q d_i\geq 2^q$, so $q\leq \log_2 d$. Hence $qmd=\wt O(md)$, and {\sc Initialize} costs $\wt O(s_C\sum_{i=1}^q \nnz(A_i)+md)$. The gain of {\sc Update} over recomputing from scratch is that the leaf sketches are linear and cached, so only $B$ is sketched, at a cost of $s_C\nnz(B)$ instead of $s_C\sum_{i=1}^q \nnz(A_i)$. We also want to point out that the sketching dimension $m$ is typically at least linear on $q$, however, it only depends on $q$, the small dimension $d$ and the precision parameter $\epsilon$, hence it is far more efficient to use this sketch representation compared to the original tensor product. Also, note that though we only design a data structure for $q$ being a power of 2, it is not hard to generalize to $q$ being an arbitrary positive integer via the technique introduced in~\cite{swyz21}. We only incur a $\log q$ factor by doing so.
\end{remark}

The approximation guarantee for our dynamic tree, which follows from \cite{akk+20} and for which we provide a brief sketch in Appendix~\ref{sec:appendix-tree_data_structure} are as follows:

\begin{theorem}[Approximation guarantee of our dynamic tree. Informal version of Theorem~\ref{thm:app:tree_approximation_guarantee}]\label{thm:tree_ds_approx}
Let $\Pi^q$ denote the sketching matrix generated by the dynamic tree. Let $M$ be any fixed matrix with $n=n_1 n_2\cdots n_q$ rows and $d'$ columns that does not depend on $\Pi^q$. If $m\geq m(d')$ with $m({\rm dim})$ from Table~\ref{tab:diff_sketch} and, for {\sc OSNAP} leaves, $s_C$ is as in the leaf sparsity paragraph above with ${\rm dim}=d'$, then with probability at least $1-\delta$, simultaneously for all $x\in \R^{d'}$,
\begin{align*}
    \|\Pi^q M x\|_2 = & ~ (1\pm\epsilon) \|M x\|_2.
\end{align*}
This applies to $M=\bigotimes_{i=1}^q A_i$ with $d'=d$, and to $M=[\bigotimes_{i=1}^q A_i,~b]$ with $d'=d+1$.
\end{theorem}

\paragraph{Choices of parameter $m$:} 

An important parameter to be chosen is the sketching dimension $m$. Intuitively, it should depend polynomially on $\epsilon^{-1}, d, q$. We list them in the following table.

\begin{table}[!ht]
    \centering
    \small
    \renewcommand{\arraystretch}{1.2}
    \begin{tabular}{|l|l|l|l|}
    \hline
     $C_{\rm base}$    & $T_{\rm base}$ &  $m({\rm dim})$ & {\bf Init time} \\ \hline
       {\sc CountSketch}  & {\sc TensorSketch} & $\epsilon^{-2} q\cdot {\rm dim}^2\cdot (1/\delta)$ & $\sum_{i=1}^q \nnz(A_i)$ \\ \hline
       {\sc OSNAP} & {\sc TensorSRHT} & $\epsilon^{-2}q^4 \cdot {\rm dim}\cdot\log(1/\delta)$ & $\wt O(q^4\epsilon^{-2}\sum_{i=1}^q \nnz(A_i))$ \\ \hline
       {\sc OSNAP} & {\sc TensorSRHT} & $\epsilon^{-2}q\,{\rm dim}^2 \log^3(1/(\epsilon\delta))$ & $\wt O(\sqrt{q}\,{\rm dim}\,\epsilon^{-1}\log(1/\delta)\sum_{i=1}^q \nnz(A_i))$ \\ \hline
    \end{tabular}
    \vspace{2mm}
    \caption{How different choices of $C_{\rm base}$ and $T_{\rm base}$ affect the sketching dimension $m({\rm dim})$ and initialization time that depends on the choice of sketch. We note that the sketching dimension $m$ is a function of ${\rm dim}$, the fundamental dimension of the problem. For example, ${\rm dim}=d$ for tensor product regression and ${\rm dim}=k$ for $k$-low rank approximation. The sketching dimension corresponds to the problem of computing an $(\epsilon, \delta, {\rm dim}, d, n)$-OSE. The last column is the leaf cost $s_C\sum_{i=1}^q \nnz(A_i)$ of {\sc Initialize}, the term $qmd$ is common to all rows, and {\sc Update} pays $s_C\nnz(B)$ for the same step.}
    \label{tab:diff_sketch}
\end{table}

\paragraph{Oblivious updates.}
The main theorems of this paper (Theorems~\ref{thm:tree_ds_approx}, \ref{thm:dynamic_k_reg}, \ref{thm:spline} and~\ref{thm:low_rank}) assume an oblivious sequence of updates and queries, meaning a sequence that is fixed in advance and independent of the random sketching matrices. Each guarantee holds at a fixed query time with probability at least $1-O(\delta)$, so a union bound over $T$ query times gives failure probability at most $O(T\delta)$.

\paragraph{Robustness against an adaptive adversary.}
An often encountered issue in dynamic algorithms is the presence of an \emph{adaptive adversary}~\cite{bkm+22}, who chooses each operation, an update $(i,B)$ or a query, after seeing the answers to all earlier queries. Such an adversary model naturally captures many applications, in which the data structure is used in an \emph{iterative process} and the input to the data structure depends on the output of the last iteration. We consider two versions. A \emph{standard} adaptive adversary sees the answers but not the random bits of the data structure. A \emph{strong} adaptive adversary also sees every random bit that the data structure has drawn so far, but not the bits it draws later. In both versions the adversary may be randomized and computationally unbounded, and a bound $T$ on the number of updates (Theorem~\ref{thm:adaptive_tree}) or of queries (Theorem~\ref{thm:adaptive_regression} and Proposition~\ref{prop:query_refresh}) is known in advance.

The tree is made robust by drawing fresh sketches on the updated path (Algorithm~\ref{alg:path_refresh}). The data structure now stores the current factors $A_1,\ldots,A_q$.

\begin{algorithm}[!ht]\caption{{\sc Update} with path refresh}\label{alg:path_refresh}
\begin{algorithmic}[1]
\Procedure{Update}{$i\in[q],B\in\R^{n_i\times d_i}$} \Comment{{\sc Initialize} is as in Algorithm~\ref{alg:tree_static} and also stores $A_1,\ldots,A_q$}
\State $A_i\gets A_i+B$
\For{$\ell=0\to\log q$}
\State $k_\ell\gets\lfloor (i-1)/2^\ell\rfloor$ \Comment{The level-$\ell$ node on the path from leaf $i-1$ to the root}
\If{$\ell=0$}
\State Choose a fresh $C_{k_0}\in\R^{m\times n_i}$ from $C_{\rm base}$, and set $J_{k_0,0}\gets C_{k_0}A_i$
\Else
\State Choose a fresh $T_{k_\ell,\ell}\in\R^{m\times m^2}$ from $T_{\rm base}$, and set $J_{k_\ell,\ell}\gets T_{k_\ell,\ell}(J_{2k_\ell,\ell-1}\otimes J_{2k_\ell+1,\ell-1})$
\EndIf
\EndFor
\EndProcedure
\end{algorithmic}
\end{algorithm}

For a node $(k,\ell)$ we write $D_{k,\ell}:=\prod_{i=k2^\ell+1}^{(k+1)2^\ell}d_i$ for the number of columns of $J_{k,\ell}$, so that $D_{k,0}=d_{k+1}$ and $D_{0,\log q}=d$.

\begin{theorem}[Adaptive guarantee for the tree]\label{thm:adaptive_tree}
Let $\epsilon\in(0,1]$, $\delta\in(0,1)$ and $T\in\mathbb{N}_+$, and set $N:=2q-1+T(1+\log q)$, $\epsilon':=\epsilon/(2(2q-1))$ and $\delta':=\delta/N$. Run {\sc DynamicTensorTree} with {\sc Initialize} as in Algorithm~\ref{alg:tree_static} and {\sc Update} as in Algorithm~\ref{alg:path_refresh}, against a strong adaptive adversary that makes at most $T$ updates. Suppose that each leaf sketch $C_k$ is drawn from an $(\epsilon',\delta',d_{k+1},n_{k+1},d_{k+1})$-OSE and each internal sketch $T_{k,\ell}$ from an $(\epsilon',\delta',D_{k,\ell},m^2,D_{k,\ell})$-OSE (Definition~\ref{def:OSE} with $\lambda=0$). Then with probability at least $1-\delta$, after {\sc Initialize} and after every update,
\begin{align*}
(1-\epsilon)\,A^\top A\preceq J_{0,\log q}^\top J_{0,\log q}\preceq(1+\epsilon)\,A^\top A ,
\end{align*}
where $A=A_1\otimes\cdots\otimes A_q$ is the current design matrix. In particular $\|J_{0,\log q}x\|_2^2=(1\pm\epsilon)\|Ax\|_2^2$ for all $x\in\R^d$. Let $s_{C,i}$ denote the sparsity of the leaf sketch that is applied to $A_i$. The hypothesis holds in the following two cases.
\begin{itemize}
\item {\sc OSNAP} leaves and {\sc TensorSRHT} internal nodes, with $s_{C,i}=\Omega(q\,d_i\log(N/\delta)/\epsilon)$ and with $m$ a power of two such that $m=\Omega(\epsilon^{-2}q^2d^2\log(N/\delta)\log^2(qdN/(\epsilon\delta)))$.
\item {\sc CountSketch} leaves and {\sc TensorSketch} internal nodes, with $m=\Omega(\epsilon^{-2}q^2d^2N/\delta)$ and $s_{C,i}=1$.
\end{itemize}
{\sc Update}$(i,B)$ takes time $\wt O(s_{C,i}(\nnz(A_i)+\nnz(B))+md)$, where $A_i$ is the factor before the update. {\sc Initialize} takes time $\wt O(\sum_{i=1}^q s_{C,i}\nnz(A_i)+qmd)$.
\end{theorem}

\begin{remark}[Scope of Theorem~\ref{thm:adaptive_tree}]\label{rem:adaptive_open}
Theorem~\ref{thm:adaptive_tree} is the guarantee of Theorem~\ref{thm:app:tree_approximation_guarantee} for $M=\bigotimes_{i=1}^qA_i$, made to hold at all times against a strong adaptive adversary. The argument goes node by node. A sketch off the updated path was drawn when its input was already fixed, and that input cannot change without the sketch being redrawn, so revealing the sketch does not hurt. The price is in the parameters. The distortions of the $2q-1$ nodes multiply, so each node needs accuracy $\Theta(\epsilon/q)$. Each internal node must embed the whole span of its children, of dimension up to $d$. The union bound runs over all $N$ draws. Compared with the third construction of Theorem~\ref{thm:app:tree_approximation_guarantee}, $m$ grows by a factor of $q$ and the factor $\log^3(1/(\epsilon\delta))$ becomes $\log(N/\delta)\log^2(qdN/(\epsilon\delta))$, and the leaf sparsity at $A_i$ becomes $\Theta(q\,d_i\log(N/\delta)/\epsilon)$ instead of $\Theta(\sqrt q\,d\log(1/\delta)/\epsilon)$. Compared with the first construction ({\sc CountSketch} leaves and {\sc TensorSketch} internal nodes), $m$ grows by a factor of $qN$, which is linear in $T$, and so does the term $md$ of the update time. The inequality of Theorem~\ref{thm:adaptive_tree} alone is easier to get without sketching. If $F_i$ has at most $d_i$ rows and $F_i^\top F_i=A_i^\top A_i$, as in Algorithm~\ref{alg:adaptive_reg}, then $W:=F_1\otimes\cdots\otimes F_q$ has at most $d$ rows and $W^\top W=A^\top A$ exactly (Lemma~\ref{lem:whitening}(i)). Maintaining $A_i^\top A_i$ and $F_i$ costs $O(\nnz(B)d_i+d_i^3)$ time per update and uses no randomness, so it is robust against any adversary. Up to constant factors this is at most the update time of Theorem~\ref{thm:adaptive_tree}, except with {\sc CountSketch} leaves when $\nnz(B)d_i$ exceeds $\nnz(A_i)+md$. What Theorem~\ref{thm:adaptive_tree} adds is that the sketch $\Pi^q$ held by the tree stays a subspace embedding for the current design matrix. Corollary~\ref{cor:adaptive_low_rank} uses exactly this. The theorem does not cover $M=[\bigotimes_{i=1}^qA_i,~b]$, because the label sketch $\Pi^qb$ passes through off-path sketches that earlier answers may have revealed, and $b$ is not a tensor product. The sketch sizes based on $\sd_\lambda$ or on $k$ do not carry over either. For low rank approximation, Theorem~\ref{thm:adaptive_tree} suffices (Corollary~\ref{cor:adaptive_low_rank}), with $m$ based on $d$ instead of $k$. Whether regression and spline regression are robust under path refresh alone is open. There every path refresh changes $\Pi^q$, so $\Pi^qb$ must be recomputed before the next query, in the time needed to compute it in Theorem~\ref{thm:app:dynamic_k_reg}, in addition to the update time $\wt O(s_C(\nnz(A_i)+\nnz(B))+md)$. We know no attack.
\end{remark}

\begin{corollary}[Low rank approximation under path refresh]\label{cor:adaptive_low_rank}
Let $\epsilon\in(0,1]$, $\delta\in(0,1)$ and $T\in\mathbb{N}_+$. Run Algorithm~\ref{alg:kronecker_low_rank} with its tree updated by Algorithm~\ref{alg:path_refresh}, and with $m$ and $s_{C,i}$ as in one of the two cases of Theorem~\ref{thm:adaptive_tree} with $\epsilon/3$ in place of $\epsilon$. Then, against a strong adaptive adversary that makes at most $T$ updates and any number of queries, with probability at least $1-\delta$ every {\sc Query} outputs a rank-$k$ matrix $C$ with $\|C-A\|_F\le\sqrt{1+\epsilon}\cdot\min_{\mathrm{rank}\text{-}k~A'}\|A'-A\|_F$ for the current $A$. {\sc Initialize} and {\sc Update} take the times of Theorem~\ref{thm:adaptive_tree}, and {\sc Query} takes $\wt O(md^{\omega-1})$ time.
\end{corollary}
\begin{proof}
On the event of Theorem~\ref{thm:adaptive_tree} with accuracy $\epsilon/3$, at every time the root is $J=\Pi^qA$ for the current sketches, and $(1-\epsilon/3)A^\top A\preceq J^\top J\preceq(1+\epsilon/3)A^\top A$. For every orthogonal projection $P$, conjugating by $I-P$ and taking traces gives $(1-\epsilon/3)\|A-AP\|_F^2\le\|J-JP\|_F^2\le(1+\epsilon/3)\|A-AP\|_F^2$. So $\Pi^q$ is a $k$-PCPSketch for $A$ with error $\epsilon/3$ and $c=0$, and the part of the correctness proof of Theorem~\ref{thm:app:low_rank} that starts from a $k$-PCPSketch with error $\epsilon/3$ applies unchanged. A {\sc Query} draws no randomness, so the number of queries does not enter.
\end{proof}

For the applications we also use two other mechanisms, stated and proved in Appendix~\ref{sec:appendix-adaptive}. For regression, Theorem~\ref{thm:adaptive_regression} keeps $K=\wt O(\sqrt{Td})$ independent copies of {\sc DTRegression}, where $T$ bounds the number of queries. It answers each query with a differentially private coordinatewise median of the answers of $\wt O(\sqrt d)$ sampled copies, following~\cite{bkm+22}. The medians are taken in coordinates in which the excess squared regression error $\|Ax-b\|_2^2-\opt^2$ is a squared Euclidean distance. These coordinates are cheap to maintain because the Gram matrix of a tensor product is the tensor product of the Gram matrices. Against a standard adaptive adversary, with probability at least $1-\beta$ every answer has cost at most $\sqrt{1+\epsilon}\cdot\opt+\xi$, for any chosen $\xi>0$, which enters the costs only through logarithms. {\sc Update} costs $K$ times the oblivious update plus $O(\nnz(B)d_i+d_i^3)$ to maintain the Gram matrices and their factors, and {\sc Query} costs $\wt O(\sqrt d)$ times the oblivious query. For all three applications, Proposition~\ref{prop:query_refresh} answers each query that follows an update from freshly drawn sketches. This is robust against a strong adaptive adversary, at the cost of one {\sc Initialize} per such query. Theorem~\ref{thm:adaptive_regression} pays $K$ initializations once and $K$ copy updates per update, whereas Proposition~\ref{prop:query_refresh}(b) pays $\kappa=O(\log(T/\delta))$ initializations at every query that follows an update. So the former is cheaper only when there are many more than $\wt O(\sqrt{Td})$ such queries, and one {\sc Initialize}, which includes computing $\Pi^qb$, costs much more than $\wt O(\sqrt{Td})$ updates of one copy. Its constants make it an asymptotic statement.

\section{Faster Dynamic Tensor Product Algorithms with Dynamic Tree}
\label{sec:applications}

In this section, we instantiate the dynamic tree data structure introduced in section~\ref{sec:tree_data_structure} for three applications: dynamic tensor product regression, dynamic tensor spline regression, and dynamic tensor low rank approximation.

\subsection{Dynamic Tensor Product Regression}\label{sec:dtpregression}
We present an algorithm (Algorithm.~\ref{alg:kronecker_reg}) for solving the dynamic Tensor product regression problem. Our algorithm uses the dynamic tree data structure to maintain a sketch of the design matrix and updates it efficiently. Additionally, we maintain a sketch of the label vector $b$ via $\Pi^q b$. During the query phase, we output the solution using the sketch of the tensor product design matrix and the sketch of the label vector. We have the following guarantees, the proof of which we defer to Appendix~\ref{sec:appendix-dtpregression}:
\begin{theorem}[Informal version of Theorem~\ref{thm:app:dynamic_k_reg}]\label{thm:dynamic_k_reg}
There is an algorithm (Algorithm~\ref{alg:kronecker_reg}) for dynamic Tensor product regression problem with the following procedures:
\begin{itemize}
    \item \textsc{Initialize}($A_1 \in \R^{n_1 \times d_1}, A_2 \in \R^{n_2 \times d_2},\dots, A_q \in \R^{n_q \times d_q}, m\in \mathbb{N}_+, C_{\mathrm{base}}, T_{\mathrm{base}}, b\in \R^{n_1\ldots n_q}$): Given matrices $A_1 \in \R^{n_1 \times d_1}, A_2 \in \R^{n_2 \times d_2}, \dots, A_q \in \R^{n_q \times d_q},$, a sketching dimension $m$, families of base sketches $C_{\mathrm{base}}$, $T_{\mathrm{base}}$ and a label vector $b\in \R^{n_1\ldots n_q}$, the data-structure pre-processes in time $\wt O(s_C\sum_{i=1}^q \nnz(A_i) + qmd)$ plus the time to compute $\Pi^q b$, which is $O(q\cdot\nnz(b))$ for {\sc CountSketch} and {\sc TensorSketch} and $\wt O(qm\cdot\nnz(b))$ for {\sc OSNAP} and {\sc TensorSRHT}.
    \item {\sc Update}$(i\in [q], B\in \R^{n_i\times d_i})$: Given a matrix $B$ and an index $i\in [q]$, the data structure updates the approximation to $A_1\otimes \ldots \otimes (A_i+B)\otimes \ldots \otimes A_q$ in time $\wt O(s_C\nnz(B) + md)$.
    \item \textsc{Query}: If $m\geq m(d+1)$, and for {\sc OSNAP} leaves also $s_C$ is as in the leaf sparsity paragraph of Section~\ref{sec:tree_data_structure} with ${\rm dim}=d+1$, where $m({\rm dim})$ is given in Table~\ref{tab:diff_sketch} and both are taken with $\epsilon$ replaced by $\epsilon/3$ (this changes $m$ and $s_C$ only by constant factors), then Query outputs an approximate solution $\hat{x}$ to the Tensor product regression problem where $\|\bigotimes_{i=1}^q A_i\hat{x} - b\|_2 = (1 \pm \epsilon) \|\bigotimes_{i=1}^q A_ix^* - b\|_2$ with probability at least $1 - \delta$ in time $O(md^{\omega-1}+d^{\omega})$ where $x^*$ is an optimal solution to the Tensor product regression problem i.e. $x^* = \arg \min_{x \in \R^d} \|\bigotimes_{i=1}^q A_ix - b\|_2$.
\end{itemize}
\end{theorem}

To instantiate Theorem~\ref{thm:dynamic_k_reg}, note that the sketch must embed $[\bigotimes_{i=1}^q A_i,~b]$, so ${\rm dim}=d+1$ in Table~\ref{tab:diff_sketch}, which changes only constants. With {\sc CountSketch} and {\sc TensorSketch}, we have $s_C=1$ and $m=\Theta(\epsilon^{-2}qd^2/\delta)$, so the update time is $\wt O(\nnz(B)+\epsilon^{-2}qd^3/\delta)$. This is input sparsity time, with a polynomial dependence on $1/\delta$. With {\sc OSNAP} and {\sc TensorSRHT} (Theorem 2 of~\cite{akk+20}), we have
\begin{align*}
    m = \wt \Theta(\epsilon^{-2} q d^2 \log^3(1/(\epsilon\delta))) \quad \text{and} \quad s_C = \Theta(\sqrt{q}\cdot d\log(1/\delta)/\epsilon),
\end{align*}
so the update time is
\begin{align*}
    \wt O(\sqrt{q}\cdot d\epsilon^{-1}\log(1/\delta)\cdot\nnz(B)+\epsilon^{-2}q d^{3}\log^3(1/(\epsilon\delta))).
\end{align*}
Thus, with the constructions of~\cite{akk+20}, leaves that run in input sparsity time come only with {\sc CountSketch} and {\sc TensorSketch}, and the constructions with a $\log(1/\delta)$ dependence multiply $\nnz(B)$ by the {\sc OSNAP} sparsity. Compared to the static algorithm~\cite{djssw19}, which can be modified for the dynamic setting, the costs are as follows. Per update, our tree costs $\wt O(s_C\nnz(B)+md)$, and each query costs $\wt O(md^{\omega-1}+d^{\omega})$. A dynamic version of~\cite{djssw19} costs $\wt O(\nnz(A_i)+\nnz(B)+\epsilon^{-2}(q+d)d/\delta)$ to resample and form the sketch after an update to $A_i$, plus $O(\epsilon^{-2}d^{\omega}/\delta)$ to solve. Over a sequence of $T$ updates, a union bound over the $T$ query times needs $\delta=\Theta(1/T)$. With {\sc CountSketch} and {\sc TensorSketch}, the sketching dimension $m$ of our tree then grows linearly in $T$, so for every $T$ the update term $md=\Theta(\epsilon^{-2}qd^3T)$ of our tree is at least a constant times the term $\epsilon^{-2}(q+d)dT$ of the baseline, because $qd^2-(q+d)=q(d^2-1)-d\geq d^2-1-d\geq 1$ when $d\geq 2$. With {\sc OSNAP} and {\sc TensorSRHT}, we have $m=\wt\Theta(\epsilon^{-2}qd^2\log^3(T/\epsilon))$ and $s_C=\Theta(\sqrt{q}\cdot d\log(T)/\epsilon)$. Comparing the $\poly(d)$ terms, the update term $md$ of our tree is $\wt\Theta(\epsilon^{-2}qd^3\log^3(T/\epsilon))$, against $\epsilon^{-2}(q+d)dT$ for the baseline. Ours is the smaller one once $T$ is at least about $qd^2\log^3(T/\epsilon)/(q+d)$, which is about $qd\log^3(T/\epsilon)$ for $d\geq q$, since then $d\leq q+d\leq 2d$. Full rounds give the same crossover for $d\geq q$: our query term $md^{\omega-1}$ is $\wt\Theta(\epsilon^{-2}qd^{\omega+1}\log^3(T/\epsilon))$, against the solve term $\epsilon^{-2}d^{\omega}T$, and the two are equal at $T\approx qd\log^3(T/\epsilon)$. So our tree is faster when $\nnz(A_i)$ dominates both $s_C\nnz(B)$ and the $\poly(d)$ terms of our tree, or, with {\sc OSNAP} and {\sc TensorSRHT}, when $s_C\nnz(B)\leq \nnz(A_i)$ and $T$ is past this crossover. For shorter sequences the baseline of~\cite{djssw19} has the smaller $\poly(d)$ term.~\cite{ffg22} improves the dependence on $d$, yielding a subquadratic update time in the form of $\wt O((\nnz(A_i)+\nnz(B)+\epsilon^{-2}q^2 d_i^\omega+\epsilon^{-1}d^{2-\theta})\log(1/\delta))$. While their algorithm is more efficient in solving the regression problem, their approach does not scale with the statistical dimension of the problem, and is geared only towards regression, whereas our data structure also handles low rank approximation.

\subsection{Dynamic Tensor Spline Regression}\label{sec:spline_regression}
Spline regression models are a well studied class of regression models~\cite{marsh2001spline}. In particular, Spline regression problems involving tensor products arise in the context of B-Splines and P-Splines~\cite{eilers1996flexible}. For a very brief but relevant introduction to Spline regression, we refer the reader to~\cite{dssw18}. In this section, we first define the Spline regression problem as it pertains to our sketching application. Algorithm~\ref{alg:kronecker_spline}, which shows how we can use our {\sc DynamicTensorTree} data structure to solve a dynamic version of the Spline regression problem is provided in Appendix~\ref{sec:appendix-splines}.

\begin{definition}[Spline Regression]
In the Spline regression problem, given a design matrix $A \in \R^{n \times d}$, a target vector $b \in \R^{n}$, a regularization matrix $L \in \R^{p \times d}$, and a non-negative penalty coefficient $\lambda \in \R$, the goal is to optimize
\begin{equation}\label{eq:spline_regression}
    \min_{x \in \R^d} \|Ax-b\|_2^2 + \lambda \cdot \|Lx\|_2^2
\end{equation}
\end{definition}

Notice that the classic ridge regression problem is a special case of spline regression (when $L = I$). In the {\sc Tensor Spline Regression} problem, we want to solve the Spline regression problem where the design matrix $A$ can be decomposed as the tensor product of multiple smaller matrices i.e. $A = A_1 \otimes \dots \otimes A_q$. Our {\sc DynamicTensorTree} data structure can be used to solve a dynamic version of {\sc Tensor Spline Regression}.  In our dynamic model, at each time step, the update is of the form $i \in [q], B \in \R^{n_i \times d_i}$, and the data structure has to update $A_i \gets A_i + B$.

Before introducing our main result, we define an important metric to measure the rank of a Spline regression:

\begin{definition}[Statistical dimension of Splines]
For matrix $A\in \R^{n\times d}$ and $L\in \R^{p\times d}$ with ${\rm rank}(L)=p$ and ${\rm rank}\left(\begin{bmatrix}
    A \\
    L
\end{bmatrix} \right)=d$. Given the generalized singular value decomposition~\cite{golubVanLoan2013Book} of $(A, L)$ such that
$
    A =  U \begin{bmatrix}
    \Sigma & 0_{p\times (d-p)} \\
    0_{(d-p)\times p} & I_{d-p}
    \end{bmatrix}R Q^\top, 
    L = V\begin{bmatrix}
        \Omega & 0_{p\times (d-p)}
    \end{bmatrix} RQ^\top,
$
where $\Sigma={\rm diag}(\sigma_1,\ldots,\sigma_p), \Omega={\rm diag}(\mu_1,\ldots,\mu_p)\in \R^{p\times p}$ are diagonal matrices. Finally, let $\gamma_i=\sigma_i/\mu_i$ for $i\in [p]$.

The statistical dimension of the Spline is defined as
$\sd_{\lambda}(A, L) =  \sum_{i=1}^p 1/(1+\lambda/\gamma_i^2)+d-p$.
\end{definition}

The statistical dimension of the Spline can be much smaller than $d$. Our data structure takes as input a number $\bar s\geq 1$ that upper bounds the statistical dimension at the query times, and we will show that the sketching dimension depends on the upper bound $\bar s$ instead of $d$. Since each term $1/(1+\lambda/\gamma_i^2)$ is at most $1$, we have $\sd_{\lambda}(A, L) = \sum_{i=1}^p 1/(1+\lambda/\gamma_i^2)+d-p\leq d$, so $\bar s=d$ is always a valid choice.

Let $x^*$ denote an optimal solution of the Spline regression problem, i.e. 
\begin{align*}
x^* = \underset{x \in \R^d}{\arg\min} \|Ax - b\|_2^2 + \lambda \cdot \|Lx\|_2^2
\end{align*}
and let $\opt \coloneqq \|Ax^* - b\|_2^2 + \lambda \cdot \|Lx^*\|_2^2$. We can compute $x^*$ given $A, b, L, \lambda$ by $x^* = (A^\top A + \lambda L^\top L)^{-1} A^{\top} b$.

\begin{theorem}[Guarantees for {\sc DTSRegression}. Informal version of Theorem~\ref{thm:app:spline}]\label{thm:spline}
Let $\epsilon\in(0,1/8]$. There exists an algorithm (Algorithm~\ref{alg:kronecker_spline}) with the following procedures:
\begin{itemize}
    \item \textsc{Initialize}$(A_1\in \R^{n_1\times d_1},\ldots,A_q\in \R^{n_q\times d_q}, \bar s\geq 1)$: Given matrices $A_1\in \R^{n_1\times d_1},\ldots,A_q\in \R^{n_q\times d_q}$ and an upper bound $\bar s\geq 1$ on the statistical dimension at the query times, the data structure chooses the sketching dimension $m$ from $\bar s$ and pre-processes in time
    \begin{align*}
    \wt O(s_C\sum_{i=1}^q \nnz(A_i)+ qmd)
    \end{align*}
    plus the time to compute $\Pi^q b$, which is $O(q\cdot\nnz(b))$ for {\sc CountSketch} and {\sc TensorSketch} and $\wt O(qm\cdot\nnz(b))$ for {\sc OSNAP} and {\sc TensorSRHT}.
    \item \textsc{Update}$(i\in [q], B\in \R^{n_i\times d_i})$: Given an index $i\in [q]$ and an update matrix $B\in \R^{n\times d}$, the data structure updates in time $\wt O(s_C\nnz(B) + md)$.
    \item \textsc{Query}: Let $A^{(t)}$ denote the tensor product $\bigotimes_{i=1}^q A_i$ of the current factors at time $t$. For every time $t$ at which $\sd_{\lambda}(A^{(t)}, L)\leq \bar s$, ${\rm rank}(L)=p$ and ${\rm rank}\left(\begin{bmatrix} A^{(t)} \\ L \end{bmatrix}\right)=d$, Query at time $t$ outputs a solution $\wt x \in \R^d$ with constant probability such the $\wt x$ is an $\epsilon$ approximate to the Tensor Spline Regression problem i.e.
    \begin{align*}
    \|A^{(t)} \wt x - b\|_2^2 + \lambda \cdot \|L \wt x\|_2^2 \leq (1+ \epsilon) \cdot \opt
    \end{align*}
    where $\opt = \min_{x \in \R^d} \|A^{(t)}x - b\|_2^2 + \lambda \cdot \|L x\|_2^2$. Query takes time $md^{(\omega - 1)} + pd^{(\omega - 1)} + d^{\omega}$.
\end{itemize}
\end{theorem}

We defer the proof of the above theorem to Appendix~\ref{sec:appendix-splines}.

The sketching dimension $m$ here depends on the upper bound $\bar s$ on the statistical dimension of the Spline, and $\bar s$ never needs to exceed $d$. One can pick {\sc OSNAP} and {\sc TensorSRHT} for approximate matrix product with accuracy $\eta=\sqrt{\epsilon/2}/\bar s$ (see the proof of Theorem~\ref{thm:app:spline}), which gives $m=\wt\Theta(\epsilon^{-1}q \bar s^2 \log^3(1/(\epsilon\delta)))$ and leaf sparsity $s_C=\Theta(\sqrt{q}\cdot\bar s\log(1/\delta)/\sqrt{\epsilon})$, so the update time is $\wt O(\sqrt{q}\cdot\bar s\epsilon^{-1/2}\log(1/\delta)\cdot\nnz(B)+\epsilon^{-1} q\bar s^2 d\log^3(1/(\epsilon\delta)))$. With {\sc CountSketch} and {\sc TensorSketch}, $s_C=1$, so the leaf step runs in input sparsity time, at the price of $m=\Theta(\epsilon^{-1}q\bar s^2/\delta)$. The improved dependence on $\epsilon^{-1}$ comes from using approximate matrix product directly, instead of OSE.
\subsection{Dynamic Tensor Low Rank Approximation}\label{sec:low_rank_approximation}
In this section, we consider a problem studied in~\cite{djssw19}: given matrices $A_1,\ldots,A_q$, the goal is to compute a low rank approximation for the tensor product matrix $A_1\otimes \ldots \otimes A_q \in \R^{n\times d}$, specifically, a rank-$k$ approximation $B\in \R^{n\times d}$ such that 
\begin{align*}
\|B-A\|_F\leq (1+\epsilon) \mathrm{OPT}_k,
\end{align*}
where $\mathrm{OPT}_k=\min_{\mathrm{rank}\text{-}k~A'}\|A'-A\|_F$ and $A=\bigotimes_{i=1}^q A_i$. The~\cite{djssw19} algorithm works as follows: they pick $q$ independent {\sc CountSketch} matrices~\cite{ccf02} with $O(\epsilon^{-2} qk^2)$ rows, then apply each sketch matrix $S_i$ to $A_i$. After that, they form the tensor product $M=\bigotimes_{i=1}^q S_i A_i$ and run SVD on $M=U\Sigma V^\top$. Finally, they output $B=AU_k^\top U_k$ in factored form (as matrices $A_1,\ldots,A_q,U_k$), where $U_k\in \R^{k\times d}$ denotes the top $k$ right singular vectors. By doing so, they achieve an overall running time of
\begin{align*} 
O(\sum_{i=1}^q \nnz(A_i)+\epsilon^{-2q}q^q k^{2q} d^{\omega-1})
\end{align*}

Two central issues around their algorithm are 1).\ exponential dependence on parameters $\epsilon^{-1}, q, k$ and 2).\ the algorithm is inherently static, so a natural way to make it dynamic is to apply $S_i$ to the update $B$, reforming the tensor product of sketches and compute the SVD.

We address these two issues simultaneously by using our tree data structure. For the sketching dimension, we observe it is enough to design a sketch that is $(\epsilon, \delta, k, d, n)$-OSE and $(\epsilon/\sqrt{k}, \delta)$-approximate matrix product, therefore, we can set sketching dimension $m=\Theta(\epsilon^{-2}q k^2/\delta)$ for {\sc CountSketch} and {\sc TensorSketch}, or $m=\wt\Theta(\epsilon^{-2}q k^2 \log^3(1/(\epsilon\delta)))$ with leaf sparsity $s_C=\Theta(\sqrt{q}\cdot k\log(1/\delta)/\epsilon)$ for {\sc OSNAP} and {\sc TensorSRHT}. As we will show later, this gives a much improved running time. By using the dynamic tree, we also provide a dynamic algorithm.

\begin{theorem}[Guarantees for {\sc LowRankMaintenance}. Informal version of Theorem~\ref{thm:app:low_rank}]\label{thm:low_rank}
There exists an algorithm (Algorithm~\ref{alg:kronecker_low_rank}) that has the following procedures
\begin{itemize}
    \item \textsc{Initialize}$(A_1\in \R^{n_1\times d_1},\ldots,A_q\in \R^{n_q\times d_q})$: Given matrices $A_1\in \R^{n_1\times d},\ldots,A_q\in \R^{n_q\times d_q}$, the data structure processes in time 
    \begin{align*}
        \wt O(s_C\sum_{i=1}^q \nnz(A_i)+qmd).
    \end{align*}
    \item \textsc{Update}$(i\in [q], B\in \R^{n_i\times d_i})$: Given an index $i\in [q]$ and an update matrix $B\in \R^{n_i\times d_i}$, the data structure updates the approximation for $A_1\otimes \ldots \otimes (A_i+B)\otimes \ldots \otimes A_q$ in time 
    \begin{align*}
    \wt O(s_C\nnz(B)+md).
    \end{align*}
    \item \textsc{Query}: Let $A$ denote the tensor product $\bigotimes_{i=1}^q A_i$. The data structure outputs a rank-$k$ approximation $C$ such that, with probability at least $1-O(\delta)$,
    \begin{align*}
        \|C-A\|_F \leq & ~ (1+\epsilon)\min_{\mathrm{rank}\text{-}k~A'}~\|A'-A\|_F.
    \end{align*}
    The time to output $C$ is $\wt O(md^{\omega-1})$.
\end{itemize}
\end{theorem}

By choosing {\sc OSNAP} and {\sc TensorSRHT} with $m=\wt\Theta(\epsilon^{-2}qk^2\log^3(1/(\epsilon\delta)))$ and $s_C=\Theta(\sqrt{q}\cdot k\log(1/\delta)/\epsilon)$, we can provide good guarantees for low rank approximation. This gives
\begin{itemize}
    \item Initialization time in $\wt O(\sqrt{q}\cdot k\epsilon^{-1}\log(1/\delta)\sum_{i=1}^q \nnz(A_i)+\epsilon^{-2}q^2 dk^2\log^3(1/(\epsilon\delta)))$, where the second term is $qmd$ as in Theorem~\ref{thm:tree_init}.
    \item Update time in $\wt O(\sqrt{q}\cdot k\epsilon^{-1}\log(1/\delta)\nnz(B)+\epsilon^{-2}qdk^2\log^3(1/(\epsilon\delta)))$.
    \item Query time in $\wt O(\epsilon^{-2}qd^{\omega-1}k^2\log^3(1/(\epsilon\delta)))$.
\end{itemize}

For {\sc CountSketch} and {\sc TensorSketch} with constant success probability, we have $m=\Theta(\epsilon^{-2}qk^2)$ and $s_C=1$, so {\sc Initialize} takes time $O(\sum_{i=1}^q \nnz(A_i))+\wt O(\epsilon^{-2}q^2dk^2)$ and {\sc Query} takes time $\wt O(\epsilon^{-2}qk^2d^{\omega-1})$. This matches the $\nnz$ term of~\cite{djssw19}, and the dependence on $q$, $k$ and $1/\epsilon$ is at most quadratic instead of their $\epsilon^{-2q}q^qk^{2q}$. We note that~\cite{djssw19} state their bound for constant $q$ and success probability $9/10$. 

\section{Conclusion \& Future Directions}\label{sec:conclusion}
In this work, we initiate the study of the Dynamic Tensor Regression, Dynamic Tensor Spline Regression, and Dynamic Tensor Low-Rank Approximation problems, and design a tree data structure {\sc DynamicTensorTree} which can be used to provide fast dynamic algorithms for these problems. The problems we study are very interesting and there are a lot of exciting future directions, some of which we discuss below. First, it is unclear whether the algorithms we proposed are optimal and so having lower bounds for the update times would be an important direction. Second, we have focused exclusively on regression problems with the $\ell_2$ norm whereas solving Dynamic Tensor Regression and Dynamic Tensor Spline Regression with respect to the more robust $\ell_1$-norm as well as for more general $p$-norms are also very appealing. Third, our main theorems assume that the sequence of updates is fixed in advance. Redrawing the sketches on the updated path keeps the tree a subspace embedding against an adaptive adversary~\cite{bkm+22}, with a larger sketching dimension and an update cost that also depends on $\nnz(A_i)$ (Theorem~\ref{thm:adaptive_tree}). For regression, private medians over $\wt O(\sqrt{Td})$ independent copies give answers of cost at most $\sqrt{1+\epsilon}\cdot\opt+\xi$, for any chosen $\xi>0$, against an adaptive adversary that does not see the random bits, where $T$ bounds the number of queries (Theorem~\ref{thm:adaptive_regression}). Redrawing the sketches on the updated path also makes low rank approximation robust, with $m$ based on $d$ instead of $k$ (Corollary~\ref{cor:adaptive_low_rank}). Whether regression and spline regression are robust under path refresh alone, with $\Pi^qb$ recomputed at each query, is an interesting open problem.

\section*{Acknowledgments}
We would like to thank all the NeurIPS 2022 reviewers of our paper and also the ICML 2022 reviewers who reviewed an earlier version of this work. Most of this work was done while LZ was at CMU. AR was supported in-part by NSF grants CCF-1652491, CCF-1934931, and CCF-1955351 during the preparation of this manuscript. LZ was supported in part by NSF grants CCF-1955217 and DMS-2022448. The corrections in this version were found and written with the help of Claude Opus 5.5. The authors have verified the corrections and take full responsibility for the content of this paper.


\ifdefined\isarxiv
 \addcontentsline{toc}{section}{References}
 \bibliographystyle{alpha}
 \bibliography{ref}
\else
 \bibliography{ref}
 \bibliographystyle{alpha}

\fi

\ifdefined\isarxiv

\else
\section*{Checklist}

\begin{enumerate}

\item For all authors...
\begin{enumerate}
  \item Do the main claims made in the abstract and introduction accurately reflect the paper's contributions and scope?
    \answerYes{} 
  \item Did you describe the limitations of your work?
    \answerYes{} Please refer to Section~\ref{sec:conclusion}.
  \item Did you discuss any potential negative societal impacts of your work?
    \answerYes{} Our work is theoretical and it provides a dynamic algorithm for tensor product-typed problems. We analyze the runtime complexity of our algorithm, which relates to energy consumption in practice. We do not foresee potential negative societal impact of our work.
  \item Have you read the ethics review guidelines and ensured that your paper conforms to them?
   \answerYes{} 
\end{enumerate}

\item If you are including theoretical results...
\begin{enumerate}
  \item Did you state the full set of assumptions of all theoretical results?
    \answerYes{} Please refer to Section~\ref{subsec:problem_formulation}.
        \item Did you include complete proofs of all theoretical results?
    \answerYes{} Please refer to Section~\ref{sec:appendix-tree_data_structure},~\ref{sec:appendix-dtpregression},~\ref{sec:appendix-splines}, and~\ref{sec:appendix-low-rank} in the supplementary materials.
\end{enumerate}

\item If you ran experiments...
\begin{enumerate}
  \item Did you include the code, data, and instructions needed to reproduce the main experimental results (either in the supplemental material or as a URL)?
    \answerNA{}
  \item Did you specify all the training details (e.g., data splits, hyperparameters, how they were chosen)?
    \answerNA{}
        \item Did you report error bars (e.g., with respect to the random seed after running experiments multiple times)?
    \answerNA{}
        \item Did you include the total amount of compute and the type of resources used (e.g., type of GPUs, internal cluster, or cloud provider)?
    \answerNA{}
\end{enumerate}

\item If you are using existing assets (e.g., code, data, models) or curating/releasing new assets...
\begin{enumerate}
  \item If your work uses existing assets, did you cite the creators?
    \answerNA{}
  \item Did you mention the license of the assets?
    \answerNA{}
  \item Did you include any new assets either in the supplemental material or as a URL?
    \answerNA{}
  \item Did you discuss whether and how consent was obtained from people whose data you're using/curating?
    \answerNA{}
  \item Did you discuss whether the data you are using/curating contains personally identifiable information or offensive content?
    \answerNA{}
\end{enumerate}

\item If you used crowdsourcing or conducted research with human subjects...
\begin{enumerate}
  \item Did you include the full text of instructions given to participants and screenshots, if applicable?
    \answerNA{}
  \item Did you describe any potential participant risks, with links to Institutional Review Board (IRB) approvals, if applicable?
   \answerNA{}
  \item Did you include the estimated hourly wage paid to participants and the total amount spent on participant compensation?
  \answerNA{}
\end{enumerate}

\end{enumerate} 
\fi
\newpage
\appendix

\section*{\centering{Appendix}}

\paragraph{Roadmap.}
In this appendix, we provide omitted details and proofs for our main paper. In particular, section~\ref{sec:appendix-prelim} of our appendix corresponds to section~\ref{sec:preliminaries} of our main paper, section~\ref{sec:appendix-tree_data_structure} to section~\ref{sec:tree_data_structure}, section~\ref{sec:appendix-dtpregression} to section~\ref{sec:dtpregression}, section~\ref{sec:appendix-splines} to section~\ref{sec:spline_regression}, and section~\ref{sec:appendix-low-rank} to section~\ref{sec:low_rank_approximation}. Section~\ref{sec:appendix-adaptive} states and proves the results for an adaptive adversary.

\section{Background} \label{sec:appendix-prelim}
In this section, we provide definitions for some terms used in the main paper and rest of the appendix.

\begin{definition}[Statistical Dimension]\label{def:SD}
For any $\lambda \geq 0$ and any positive semidefinite matrix $M \in \R^{n \times n}$, the $\lambda$-statistical dimension of $M$ is
\begin{align*}
    \sd_{\lambda}(M) &\coloneqq \tr[ M(M+\lambda I_n)^{-1} ]
\end{align*}
\end{definition}

\begin{definition}[Oblivious Subspace Embedding~\cite{s06,akk+20}]\label{def:OSE}
Let $\epsilon,\delta\in (0,1), \mu > 0,$ and $d,n\geq 1$ be integers. An $(\epsilon,\delta,\mu,d,n)$-Oblivious Subspace Embedding (OSE) is a distribution over $m\times d$ matrices $\Pi$ with the guarantee that for any fixed $\lambda\geq 0$ and any fixed matrix $A\in \R^{d\times n}$ with $\sd_{\lambda}(A^\top A) \leq \mu$, with probability at least $1-\delta$,
\begin{align*}
(1-\epsilon)(A^\top A+\lambda I_n) \preceq (\Pi A)^\top (\Pi A)+\lambda I_n \preceq (1+\epsilon)(A^\top A+\lambda I_n).
\end{align*}
For $\lambda=0$, the inverse in Definition~\ref{def:SD} is read as the pseudo-inverse, so that $\sd_0(A^\top A)=\rank(A)$.
\end{definition}

\begin{definition}[Approximate Matrix Product]\label{def:amp}
Given $\epsilon,\delta>0$, we say that a distribution ${\cal D}$ over $m\times d$ matrices $\Pi$ has the $(\epsilon,\delta)$-approximate matrix product property if for every $C,D\in \R^{d\times n}$, with probability at least $1-\delta$,
\begin{align*}
    \|C^\top \Pi^\top \Pi D-C^\top D\|_F \leq & ~ \epsilon \|C\|_F\|D\|_F.
\end{align*}
\end{definition}


\begin{definition}[\textsc{OSNAP} matrix~\cite{nn13}]
\label{def:osnap}
For every sparsity parameter $s$, target dimension $m$, and positive integer $d$, the OSNAP transform with sparsity $s$ is defined as
\begin{align*}
    S_{r,j} = & ~ \frac{1}{\sqrt s} \cdot \delta_{r,j}\cdot \sigma_{r,j},
\end{align*}
for all $r\in [m]$, $j\in [d]$, where $\sigma_{r,j}\in \{\pm 1\}$ are independent and uniform Rademacher random variables and $\delta_{r,j}$ are Bernoulli random variables satisfying
\begin{itemize}
    \item For every $i\in [d]$, $\sum_{r\in [m]}\delta_{r,i}=s$. In other words, every column has exactly $s$ nonzero entries.
    \item For all $r\in [m]$ and all $i\in [d]$, $\E[\delta_{r,i}]=s/m$.
    \item The $\delta_{r,i}$'s are negatively correlated: $\forall\ T\subset [m]\times [d],\ \E[\prod_{(r,i)\in T}\delta_{r,i}]\leq \prod_{(r,i)\in T} \E[\delta_{r,i}]=(\frac{s}{m})^{|T|}$.
\end{itemize}
\end{definition}

\begin{definition}[\textsc{TensorSRHT}~\cite{akk+20}]
\label{def:TensorSRHT}
We define the \textsc{TensorSRHT} $S : \R^d \times \R^d \rightarrow \R^m$ as $S = \frac{1}{\sqrt m} P \cdot (H D_1\times H D_2)$, where each row of $P\in \{0,1\}^{m\times d}$ contains only one $1$ at a random coordinate, one can view $P$ as a sampling matrix. $H$ is a $d \times d$ Hadamard matrix, and $D_1, D_2$ are two $d\times d$ independent diagonal matrices with diagonals that are each independently set to be a Rademacher random variable (uniform in $\{-1,1\}$). 
\end{definition}

\begin{definition}[\textsc{CountSketch} matrix~\cite{ccf02}]\label{def:cs}
Let $h : [n] \rightarrow [b]$  be a random $2$-wise independent hash function and $\sigma : [n] \rightarrow \{-1,+1\}$ be a random $4$-wise independent hash function. Then $R\in\R^{b\times n}$ is a \textsc{CountSketch} matrix if we set $R_{h(i),i} = \sigma(i)$ for all $i\in[n]$ and all the other entries of $R$ to zero.
\end{definition}

\begin{definition}[\textsc{TensorSketch} matrix~\cite{p13,pp13}]\label{def:TensorSketch}
Let $h_1,h_2:[m] \rightarrow [s]$ be 3-wise independent hash functions, also let $\sigma:[m] \rightarrow \{\pm 1\}$ be a 4-wise independent random sign function. The degree two \textsc{TensorSketch} transform, $S:\R^m\times \R^m\rightarrow \R^s$ is defined as follows: for any $i,j\in [m]$ and $r\in [s]$,
\begin{align*}
S_{r,(i,j)} = & ~ \sigma_1(i)\cdot \sigma_2(j)\cdot \mathbf{1}[h_1(i)+h_2(j)=r~\textnormal{mod}~s].
\end{align*}
\end{definition}

\begin{definition}[Subsampled Randomized Hadamard Transform ({\sc SRHT}) matrix \cite{ldfu13}]\label{def:srht}
We say $R \in \R^{b\times n}$ is a subsampled randomized Hadamard transform matrix\footnote{In this case, we require $\log{n}$ to be an integer.} if it is of the form $R = \sqrt{n/b} \cdot SHD$, where $S \in \R^{b \times n}$ is a random matrix whose rows are $b$ uniform samples (without replacement) from the standard basis of $\R^n$, $H \in \R^{n \times n}$ is a normalized Walsh-Hadamard matrix, and $D \in \R^{n \times n}$ is a diagonal matrix whose diagonal elements are i.i.d. Rademacher random variables.
\end{definition}

\section{Proofs for Dynamic Tree Data Structure (section ~\ref{sec:tree_data_structure})}\label{sec:appendix-tree_data_structure}
In this section, we provide proofs for the running time and approximation guarantees of our dynamic tree data structure (Algorithm~\ref{alg:tree_static}).

\subsection{Running Time of Algorithm~\ref{alg:tree_static}}
\begin{theorem}[Formal version of Theorem~\ref{thm:tree_init}]\label{thm:app:tree_init}
Let $s_C$ denote the maximum number of nonzero entries in a column of a leaf sketch $C_k$, so that applying $C_k$ to a matrix $X$ costs $O(s_C\nnz(X))$. We have $s_C=1$ for {\sc CountSketch}. For {\sc OSNAP} leaves with {\sc TensorSRHT} internal nodes (Lemma~\ref{lem:osnap_tensorsrht}), $s_C=\Theta(\sqrt{q}\cdot{\rm dim}\cdot\log(1/\delta)/\epsilon)$ for an OSE of a subspace of dimension ${\rm dim}$ (see Table~\ref{tab:diff_sketch}), and $s_C=\Theta(\sqrt{q}\log(1/\delta)/\epsilon')$ for approximate matrix product with accuracy $\epsilon'$. For the construction of Lemma~\ref{lem:srht_tensorsrht}, $s_C=\wt O(q^4/\epsilon^2)$. There is an algorithm (Algorithm~\ref{alg:tree_static}) that has the following procedures:
\begin{itemize}
    \item \textsc{Initialize}($A_1 \in \R^{n_1 \times d_1}, A_2 \in \R^{n_2 \times d_2},\dots, A_q \in \R^{n_q \times d_q}, m\in \mathbb{N}_+, C_{\mathrm{base}}, T_{\mathrm{base}}$): Given matrices $A_1 \in \R^{n_1 \times d_1}, A_2 \in \R^{n_2 \times d_2}, \dots, A_q \in \R^{n_q \times d_q}$, a sketching dimension $m$, families of base sketches $C_{\mathrm{base}}$ and $T_{\mathrm{base}}$, the data-structure pre-processes in time $\wt O(s_C\sum_{i=1}^q \nnz(A_i) + qmd)$ where $d = d_1 d_2 \dots d_q$.
    \item {\sc Update}$(i\in [q], B\in \R^{n_i\times d_i})$: Given a matrix $B$ and an index $i\in [q]$, the data structure updates the approximation to $A_1\otimes \ldots \otimes (A_i+B)\otimes \ldots \otimes A_q$ in time $\wt O(s_C\nnz(B) + md)$.
\end{itemize}
\end{theorem}
Note: We will assume that $n_1 = n_2 = \dots = n_q$ and $d_1 = d_2 = \dots = d_q$ for the following computations. To modify our {\sc DynamicTensorTree} to work for the non-uniform case, we only need to choose the base sketches at the leaves of the appropriate dimension and also have different dimensions for the internal nodes.
\begin{proof}
{\bf Proof of \textsc{Initialization}}. The initialization part has $2$ key steps - applying the $T_{\mathrm{base}}$ sketches to the leaves and applying $S_{\mathrm{base}}$ sketches to each of the internal nodes.

\begin{itemize}
    \item Applying a leaf sketch $C_i\in\R^{m \times n_i}$ to a matrix $A_i\in \R^{n_i \times d_i}$ takes time $O(s_C\nnz(A_i))$, because each nonzero entry of $A_i$ meets the $s_C$ nonzero entries of one column of $C_i$, plus $O(md_i)$ time to write the output, which is at most the $qmd$ term below. For {\sc CountSketch}, $s_C=1$, so this is input sparsity time $O(\nnz(A_i))$. Therefore, the total time for this step is $O(s_C\sum_{i = 1}^q \nnz(A_i))$.
    \item For all internal nodes, let us consider the time of computing one node at level $\ell \in [\log_2 q]$ using {\sc TensorSRHT}. Consider the computation of $J_{k,\ell} \in \R^{m\times d_1^{2^\ell}}$, we pick a {\sc TensorSRHT} $T_{k,\ell}\in \R^{m\times m^2}$ and apply it to $J_{2k,\ell-1}\otimes J_{2k+1,\ell-1}$. This can be viewed as computing the tensor product of column vectors of $J_{2k,\ell-1}$ and $J_{2k+1,\ell-1}$: each column of $J_{k,\ell}$ is a tensor product of two columns of its two children after applying $T_{k,\ell}$. Each column takes time $O(m \log m)$ and the total number of columns of $J_{k,\ell}$ is $d_1^{2^\ell}$. Hence, the time of forming $J_{k,\ell}$ is $O(md_1^{2^\ell}\log m)$.
    \item For each level, there are $q/2^\ell$ nodes, so $O(q/2^\ell md_1^{2^\ell})$ time in total for level $\ell$.
    \item Sum over all levels for $\ell\in [\log q]$, we have
    \begin{align*}
        m \log m (\sum_{\ell=1}^{\log q} \frac{q}{2^\ell} d_1^{2^\ell}) \leq & ~ m \log m (\sum_{\ell=1}^{\log q} \frac{q}{2^\ell} d_1^{q}) \\
        = & ~ md\log m(\sum_{\ell=1}^{\log q} \frac{q}{2^\ell}) \\
        \leq & ~ 2q md\cdot\log m.
    \end{align*}

    \item Thus, the total time for initialization is 
    \begin{align*}
        \wt O(s_C\sum_{i=1}^q \nnz(A_i) + qmd).
    \end{align*}
      
\end{itemize}
{\bf Proof of \textsc{Update}}. We note that \textsc{Update} can be decomposed into the following parts:
\begin{itemize}
    \item Apply the leaf sketch $C_i$ to the update matrix $B$, which takes $O(s_C\nnz(B))$ time by the same argument as in \textsc{Initialization}, and $s_C=1$ for {\sc CountSketch}.
    \item Updating the leaf node takes $O(md_i)$ time because $C_i B \in \R^{m\times d_i}$.
    \item For each level $\ell\in [\log_2 q]$, we first compute the update tree by applying \textsc{TensorSketch} $T_{k,\ell} \in \R^{m \times m^2}$ to $J_{2k,\ell-1} \otimes J_{2k+1,\ell-1}$ which takes time $O(md_1^{2^\ell} \log m)$.
    \item We also need to update \textbf{one} internal node for each level, which takes $\wt O(md_1^{2^{\ell}})$ time. This is at most $O(md \log m)$ for the root. Since there are $\log q$ levels, the total time for updating all internal nodes is also $\wt O(md)$. 
\end{itemize}
Hence, the total runtime of \textsc{Update} is
\begin{align*}
    & ~ \wt O(s_C\nnz(B)+ m d_1 + md ) \\
    = & ~ \wt O(s_C\nnz(B) + md).
\end{align*}
\end{proof}

\subsection{Approximation Guarantee of Algorithm~\ref{alg:tree_static}}
In this section, we provide the approximation guarantee of Algorithm~\ref{alg:tree_static}.

\begin{theorem}\label{thm:app:tree_approximation_guarantee}
Let $A_1\in \R^{n_1\times d_1},\ldots,A_q\in \R^{n_q\times d_q}$ be the input matrices, $m\in \mathbb{N}_+$ be the sketching dimension, let $C_{\mathrm{base}}, T_{\mathrm{base}}$ be appropriate family of sketching matrices. Let $\Pi^q$ be the sketching matrix generated by Algorithm~\ref{alg:tree_static}. Let $M\in \R^{n\times d'}$, where $n=n_1 n_2\cdots n_q$, be any fixed matrix that does not depend on $\Pi^q$, for example $M=\bigotimes_{i=1}^q A_i$ with $d'=d$, or $M=[\bigotimes_{i=1}^q A_i,~b]$ with $d'=d+1$. Then with probability at least $1-\delta$, simultaneously for all $x\in \R^{d'}$,
\begin{align*}
    \|\Pi^q M x\|_2 = & ~ (1\pm\epsilon) \|M x\|_2,
\end{align*}
if
\begin{itemize}
    \item $C_{\mathrm{base}}$ is {\sc CountSketch}, $T_{\mathrm{base}}$ is {\sc TensorSketch} and $m=\Omega(\epsilon^{-2}qd'^2 1/\delta)$.
    \item $C_{\mathrm{base}}$ is {\sc OSNAP}, $T_{\mathrm{base}}$ is {\sc TensorSRHT}, $m=\wt \Omega(\epsilon^{-2}q^4 d'\log(1/\delta))$, and the leaf sparsity is $s_C=\wt O(q^4/\epsilon^2)$.
    \item $C_{\mathrm{base}}$ is {\sc OSNAP}, $T_{\mathrm{base}}$ is {\sc TensorSRHT}, $m=\wt \Omega(\epsilon^{-2}qd'^2 \log^3(1/(\epsilon\delta)))$, and the leaf sparsity is $s_C=\Theta(\sqrt{q}\cdot d'\log(1/\delta)/\epsilon)$.
\end{itemize}
\end{theorem}

First we will discuss the theorems of \cite{akk+20} which lead to our approximation guarantees and then discuss the specifics for each choice of base sketches seperately. 

\begin{lemma}[Theorem 1 of~\cite{akk+20} and its proof]\label{lem:cs_tensorsketch}
For every positive integers $n,q,d$, every $\epsilon,\sd_\lambda>0$ and every $\delta\in(0,1)$, there exists a distribution on linear sketches $\Pi^q\in \R^{m\times d^q}$ such that: 1).\ If $m=\Omega(\epsilon^{-2}q\sd_\lambda^2/\delta)$, then $\Pi^q$ is an $(\epsilon,\delta,\sd_\lambda,d^q,n)$-OSE (Def.~\ref{def:OSE}). 2).\ If $m=\Omega(\epsilon^{-2}q/\delta)$, then $\Pi^q$ has the $(\epsilon,\delta)$-approximate matrix product (Def.~\ref{def:amp}).
\end{lemma}

\begin{lemma}[Theorem 2 of~\cite{akk+20}]\label{lem:osnap_tensorsrht}
For every positive integers $n,q,d$, every $\epsilon,\sd_\lambda>0$ and every $\delta\in(0,1)$, there exists a distribution on linear sketches $\Pi^q\in \R^{m\times d^q}$, formed by {\sc OSNAP} leaves (Def.~\ref{def:osnap}) with $s_C\leq m$ nonzero entries per column and {\sc TensorSRHT} internal nodes, such that: 1).\ If $m=\wt \Omega(\epsilon^{-2}q\sd_\lambda^2\log^3(1/(\epsilon\delta)))$ and $s_C=\Omega(\sqrt{q}\cdot\sd_\lambda\log(1/\delta)/\epsilon)$, then $\Pi^q$ is an $(\epsilon,\delta,\sd_\lambda,d^q,n)$-OSE (Def.~\ref{def:OSE}). 2).\ If $m=\wt \Omega(\epsilon^{-2}q\log^3(1/(\epsilon\delta)))$ and $s_C=\Omega(\sqrt{q}\log(1/\delta)/\epsilon)$, then $\Pi^q$ has the $(\epsilon,\delta)$-approximate matrix product (Def.~\ref{def:amp}).
\end{lemma}

\begin{lemma}[Theorem 3 of~\cite{akk+20}]\label{lem:srht_tensorsrht}
For every positive integers $n,q,d$, every $\epsilon,\sd_\lambda>0$, there exists a distribution on linear sketches $\Pi^q\in \R^{m\times d^q}$, formed by {\sc OSNAP} leaves (Def.~\ref{def:osnap}) with $s_C=\wt O(q^4/\epsilon^2)$ nonzero entries per column and {\sc TensorSRHT} internal nodes, which is an $(\epsilon,1/\poly(n),\sd_\lambda,d^q,n)$-OSE (Def.~\ref{def:OSE}), provided that the integer $m$ satisfies $m = \wt \Omega(q^4 \sd_{\lambda}/\epsilon^2)$
\end{lemma}

The three results of~\cite{akk+20} are stronger than what we need here, because their guarantees hold for every fixed matrix of the proper size, not only for tensor products, each with probability at least $1-\delta$. Also, when the statistical dimension $\sd_\lambda$ is smaller than the rank of the design matrix, the sketching dimension is even smaller, which is important in our Spline regression application. 

In each case, we apply the lemma with $\lambda=0$ to $M$, which has $\sd_0(M^\top M)=\rank(M)\leq d'$. By Definition~\ref{def:OSE}, with probability at least $1-\delta$,
\begin{align*}
    (1-\epsilon) M^\top M \preceq (\Pi^q M)^\top \Pi^q M \preceq (1+\epsilon) M^\top M.
\end{align*}
On this event, evaluating the three quadratic forms at any $x\in \R^{d'}$ gives $(1-\epsilon)\|Mx\|_2^2 \leq \|\Pi^q M x\|_2^2 \leq (1+\epsilon)\|Mx\|_2^2$ for all $x\in \R^{d'}$. Taking square roots, and using $\sqrt{1-\epsilon}\geq 1-\epsilon$ and $\sqrt{1+\epsilon}\leq 1+\epsilon$ for $\epsilon\in(0,1)$, we obtain $\|\Pi^q M x\|_2=(1\pm\epsilon)\|Mx\|_2$ simultaneously for all $x\in \R^{d'}$.

We then proceed to prove the individual guarantees for different choice of base sketches item by item.
\begin{itemize}
    \item By Lemma~\ref{lem:cs_tensorsketch}, we know that in order for $\Pi^q$ to be an OSE, we will require $m=\Omega(\epsilon^{-2}qd'^2 1/\delta)$. Note that $\sd_0(M^\top M)=\rank(M)\leq d'$ in this case. We point out that the dependence on success probability is $1/\delta$, this means that we can not obtain a high probability version. This is the main reason that Theorem 1 of~\cite{akk+20} is stated only for the constant failure probability $1/10$.
    \item By Lemma~\ref{lem:srht_tensorsrht}, we need $m=\wt \Omega(\epsilon^{-2}q^4 d'\log(1/\delta))$, with leaf sparsity $s_C=\wt O(q^4/\epsilon^2)$. We get a nearly linear dependence on $d$ in the expense of worse dependence on $q$. 
    \item By Lemma~\ref{lem:osnap_tensorsrht}, we will require $m=\wt \Omega(\epsilon^{-2}qd'^2 \log^3(1/(\epsilon\delta)))$ and leaf sparsity $s_C=\Theta(\sqrt{q}\cdot d'\log(1/\delta)/\epsilon)$. In this scenario, the sketching dimension depends polylogarithmically on $1/\delta$, hence we can aim for a high probability guarantee for our OSE, but the leaf cost carries the factor $s_C$.
\end{itemize}

\section{Proofs for Dynamic Tensor Product Regression (section \ref{sec:dtpregression})}
\label{sec:appendix-dtpregression}

\begin{algorithm}[!ht]\caption{Dynamic Tensor Product Regression}
\label{alg:kronecker_reg}
\begin{algorithmic}[1]
\State {\bf data structure} {\sc DTRegression}
\State
\State {\bf members}
\State \hspace{4mm} $\textsc{DynamicTensorTree DTT} $
\State \hspace{4mm} $A_1\in \R^{n_1\times d_1},\ldots,A_q\in \R^{n_q\times d_q}$
\State \hspace{4mm} $\wt b\in \R^{m}$
\State {\bf end members}
\State
\Procedure{Initialize}{$A_1\in \R^{n_1\times d_1},\ldots,A_q\in \R^{n_q\times d_q}, m,C_{\mathrm{base}},T_{\mathrm{base}}, b\in \R^n$}
\State $\textsc{DTT.Initialize}(A_1,\ldots,A_q, m,C_{\mathrm{base}},T_{\mathrm{base}})$
\State $A_1\gets A_1,\ldots,A_q\gets A_q$
\State Let $\Pi_q \in \R^{m\times n}$ be the sketching matrix corresponds to {\sc DTT}
\State $\wt b\gets \Pi_q b$
\EndProcedure
\State
\Procedure{Update}{$i\in [q], B\in \R^{n_i\times d_i}$}
\State $\textsc{DTT.Update}(i, B)$
\State $A_i\gets A_i+B$
\EndProcedure
\State
\Procedure{Query}{$ $}
\State $M\gets \textsc{DTT.}J_{0,\log q}$
\State $\wt x \gets \arg \min_{x \in \R^d} \|Mx - \wt b\|_2^2$
\State \Return $\wt x$
\EndProcedure
\end{algorithmic}
\end{algorithm}

\begin{theorem}[Formal version of Theorem~\ref{thm:dynamic_k_reg}]\label{thm:app:dynamic_k_reg}
There is a dynamic data structure for dynamic Tensor product regression problem with the following procedures:
\begin{itemize}
    \item \textsc{Initialize}($A_1 \in \R^{n_1 \times d_1}, A_2 \in \R^{n_2 \times d_2},\dots, A_q \in \R^{n_q \times d_q}, m\in \mathbb{N}_+, C_{\mathrm{base}}, T_{\mathrm{base}}, b\in \R^{n_1\ldots n_q}$): Given matrices $A_1 \in \R^{n_1 \times d_1}, A_2 \in \R^{n_2 \times d_2}, \dots, A_q \in \R^{n_q \times d_q},$, a sketching dimension $m$, families of base sketches $C_{\mathrm{base}}$, $T_{\mathrm{base}}$ and a label vector $b\in \R^{n_1\ldots n_q}$, the data-structure pre-processes in time $\wt O(s_C\sum_{i=1}^q \nnz(A_i) + qmd)$ plus the time to compute $\Pi^q b$, where $s_C$ is the leaf sparsity of Theorem~\ref{thm:app:tree_init}. The time to compute $\Pi^q b$ is $O(q\cdot\nnz(b))$ when $C_{\mathrm{base}}$ is {\sc CountSketch} and $T_{\mathrm{base}}$ is {\sc TensorSketch}, and $\wt O(qm\cdot\nnz(b))$ when $C_{\mathrm{base}}$ is {\sc OSNAP} and $T_{\mathrm{base}}$ is {\sc TensorSRHT}.
    \item {\sc Update}$(i\in [q], B\in \R^{n_i\times d_i})$: Given a matrix $B$ and an index $i\in [q]$, the data structure updates the approximation to $A_1\otimes \ldots \otimes (A_i+B)\otimes \ldots \otimes A_q$ in time $\wt O(s_C\nnz(B) + md)$.
    \item \textsc{Query}: If $m$, and for {\sc OSNAP} leaves also the leaf sparsity $s_C$, satisfy the conditions of Theorem~\ref{thm:app:tree_approximation_guarantee} with $d'=d+1$ and with $\epsilon$ replaced by $\epsilon/3$ (this changes $m$ and $s_C$ only by constant factors), then Query outputs an approximate solution $\hat{x}$ to the Tensor product regression problem where $\|\bigotimes_{i=1}^q A_i\hat{x} - b\|_2 = (1 \pm \epsilon) \|\bigotimes_{i=1}^q A_ix^* - b\|_2$ with probability at least $1 - \delta$ in time $O(md^{\omega-1}+d^{\omega})$ where $x^*$ is an optimal solution to the Tensor product regression problem i.e. $x^* = \arg \min_{x \in \R^d} \|\bigotimes_{i=1}^q A_ix - b\|_2$.
\end{itemize}
\end{theorem}

\begin{proof}
{\bf Proof of \textsc{Initialization}}. The initialization part has $3$ steps:
\begin{itemize}
    \item Initializing DTT takes time $\wt O(s_C\sum_{i=1}^q \nnz(A_i) + qmd)$ from Theorem~\ref{thm:app:tree_init}.
    \item Initializing all the matrices $A_i, i \in [q]$ takes time $\wt O(\sum_{i=1}^q \nnz(A_i))$.
    \item Computing $\wt b=\Pi^q b$. Write
    \begin{align*}
        b = \sum_{j} b_j\, e_{j_1}\otimes \cdots \otimes e_{j_q},
    \end{align*}
    where the sum runs over the $\nnz(b)$ indices $j=(j_1,\ldots,j_q)$ with $b_j\neq 0$, and $e_{j_i}\in \R^{n_i}$ is a standard basis vector. By linearity, $\Pi^q b=\sum_j b_j\, \Pi^q(e_{j_1}\otimes \cdots \otimes e_{j_q})$. By the mixed-product property, $\Pi^q(e_{j_1}\otimes \cdots \otimes e_{j_q})$ is the root of the tree computed from the vectors $e_{j_1},\ldots,e_{j_q}$ in place of the matrices $A_1,\ldots,A_q$: the leaf $k$ computes $C_k e_{j_{k+1}}$, and each internal node applies $T_{k,\ell}$ to the tensor product of its two children. So each term takes one leaf-to-root pass through the $q$ leaves and the $q-1$ internal nodes, and the $\nnz(b)$ results are summed at the root, starting from the zero vector in $\R^m$ in $O(m)$ time, which is at most the $qmd$ term.
    \begin{itemize}
        \item If $C_{\mathrm{base}}$ is {\sc CountSketch} and $T_{\mathrm{base}}$ is {\sc TensorSketch}, every intermediate vector has one nonzero entry. A {\sc CountSketch} maps $e_a$ to $\sigma(a)e_{h(a)}$ (Def.~\ref{def:cs}), and a {\sc TensorSketch} maps $e_a\otimes e_{a'}$ to $\sigma_1(a)\sigma_2(a')e_{r}$ with $r=(h_1(a)+h_2(a'))~\textnormal{mod}~s$ (Def.~\ref{def:TensorSketch}). So each node takes $O(1)$ time, a pass costs $O(q)$, and adding its result to the root sum costs $O(1)$. The total time is $O(q\cdot\nnz(b))$.
        \item If $C_{\mathrm{base}}$ is {\sc OSNAP} and $T_{\mathrm{base}}$ is {\sc TensorSRHT}, the output $C_k e_{j_{k+1}}$ of a leaf is a column of $C_k$, so it has $s_C\leq m$ nonzero entries and takes $O(s_C)$ time. A {\sc TensorSRHT} node with inputs $u,v\in \R^m$ computes $HD_1u$ and $HD_2v$ in $O(m\log m)$ time and then $m$ sampled entries of their tensor product in $O(m)$ time (Def.~\ref{def:TensorSRHT}), and its output is a dense vector. So a pass costs $O(qs_C+(q-1)m\log m)=\wt O(qm)$, and adding its result to the root sum costs $O(m)$. The total time is $\wt O(qm\cdot\nnz(b))$.
    \end{itemize}
\end{itemize}
Therefore, since $s_C\geq 1$, the total time for initialization is
\begin{align*}
& ~ \wt O(s_C\sum_{i=1}^q \nnz(A_i) + qmd + \sum_{i=1}^q \nnz(A_i))\\
= & ~ \wt O(s_C\sum_{i=1}^q \nnz(A_i) + qmd)
\end{align*}
plus the time to compute $\Pi^q b$, which is $O(q\cdot\nnz(b))$ for {\sc CountSketch} and {\sc TensorSketch}, and $\wt O(qm\cdot\nnz(b))$ for {\sc OSNAP} and {\sc TensorSRHT}.

{\bf Proof of \textsc{Update}}. We just update our DTT data structure. This takes time $\wt O(s_C\nnz(B) + md)$ from Theorem~\ref{thm:app:tree_init}.

{\bf Proof of \textsc{Query} time}. We can view the Query part as having the following $5$ steps:

\begin{itemize}
    \item Initializing $M$ takes time $\wt O(md)$ since $M \in \R^{m \times d}$.
    \item $M^\top (\wt b)$, this takes $O(md)$ time because we are multiplying a $d \times m$ matrix with a $m \times 1$ vector.
    \item $M^\top \cdot M$, this step takes $\Tmat(d,m,d)$ time as we are multiplying an $d \times m$ matrix with an $m \times d$ matrix.
    \item $(M^\top M)^{-1}$ this step takes $\Tmat(d,d,d)$ time as it is inverting $M^\top M$ which is of size $d \times d$.
    \item $(M^\top M)^{-1} (M^\top \wt b)$. In this step, we are multiplying $(M^\top M)^{-1}$ which is a $d \times d$ matrix with a vector $M^\top \wt b$ of size $d$. This takes time $O(d^2)$.
\end{itemize}
Overall, the total running time of $\textsc{Query}$ is
\begin{align*}
    & ~ \wt O(md + md + \Tmat(d,m,d) +  \Tmat(d,d,d) + d^2) \\
    = & ~ \wt O(md + md^{(\omega - 1)} + d^{\omega} + d^2) \\
    = & ~ \wt O(md^{(\omega - 1)} + d^{\omega})
\end{align*}

{\bf Proof of Correctness for \textsc{Query}}. As in the proof of Theorem 3.1 in~\cite{dssw18}, we use a subspace embedding for the column span of $[A,~b]$. Write $A=\bigotimes_{i=1}^q A_i$ for the current factors, and let $\Pi^q$ be the sketching matrix formed by the sketches $C_k$ and $T_{k,\ell}$ chosen in \textsc{Initialize}.

First, the root stores $\Pi^q A$ after any sequence of updates. By the mixed-product property $(XU)\otimes (YV)=(X\otimes Y)(U\otimes V)$, the root matrix computed by \textsc{Initialize} is $\Pi^q A$, where $\Pi^q$ depends only on the sketches and not on the factors. Every node's sketch is linear in the input that changes: $C_k(A_{k+1}+B)=C_kA_{k+1}+C_kB$ at a leaf, and $T_{k,\ell}((X+\Delta X)\otimes Y)=T_{k,\ell}(X\otimes Y)+T_{k,\ell}(\Delta X\otimes Y)$ at an internal node whose left child changes (and in the same way when the right child changes). An update of $A_i$ changes only the leaf $i-1$ and its ancestors, and by these identities \textsc{Update} adds to each of them exactly the change of its value. So, by induction on the number of updates, every node stores the matrix that \textsc{Initialize} would compute from the current factors with the same sketches. In particular $J_{0,\log q}=\Pi^q A$. The vector $\wt b=\Pi^q b$ computed in \textsc{Initialize} stays valid because $b$ and the sketches never change. Hence \textsc{Query} returns $\wt x \in \arg\min_{x\in \R^d}\|\Pi^q(Ax-b)\|_2$, which is the $\hat x$ of the statement.

Next, the matrix $M=[A,~b]\in \R^{n\times (d+1)}$ is fixed independently of $\Pi^q$, because the sequence of updates is fixed in advance. We apply Theorem~\ref{thm:app:tree_approximation_guarantee} to $M$ with $d'=d+1$ and accuracy $\epsilon/3$, which is allowed by the assumption on $m$ and $s_C$ in the statement. With probability at least $1-\delta$, for all $y\in \R^{d+1}$ we have $\|\Pi^q M y\|_2=(1\pm\epsilon/3)\|My\|_2$. For $x\in \R^d$, taking $y=(x,-1)$ gives $My=Ax-b$, so
\begin{align*}
    \|\Pi^q(Ax-b)\|_2 = & ~ (1\pm\epsilon/3)\|Ax-b\|_2 \quad \text{for all } x\in \R^d.
\end{align*}
This event covers all $x$ at once, which is needed because $\wt x$ depends on $\Pi^q$. On this event,
\begin{align*}
    \|A\wt x-b\|_2 \leq & ~ \frac{1}{1-\epsilon/3}\|\Pi^q(A\wt x-b)\|_2 \\
    \leq & ~ \frac{1}{1-\epsilon/3}\|\Pi^q(Ax^*-b)\|_2 \\
    \leq & ~ \frac{1+\epsilon/3}{1-\epsilon/3}\|Ax^*-b\|_2 \\
    \leq & ~ (1+\epsilon)\|Ax^*-b\|_2.
\end{align*}
The first and third steps use the previous display at $\wt x$ and at $x^*$, and the second step uses that $\wt x$ minimizes $\|\Pi^q(Ax-b)\|_2$. The last step uses $\frac{1+a}{1-a}\leq 1+3a$ for $a\in[0,1/3]$, applied with $a=\epsilon/3$, which lies in $[0,1/3]$ since $\epsilon\leq 1$. Multiplying by $1-a>0$, this inequality is equivalent to $1+a\leq (1+3a)(1-a)=1+2a-3a^2$, that is, to $a-3a^2=a(1-3a)\geq 0$, which holds for $a\in[0,1/3]$. Finally, $\|A\wt x-b\|_2\geq \|Ax^*-b\|_2$ by the optimality of $x^*$. Hence $\|A\wt x-b\|_2=(1\pm\epsilon)\|Ax^*-b\|_2$ with probability at least $1-\delta$.
\end{proof}

\section{Proofs for the adaptive adversary}\label{sec:appendix-adaptive}

In this section we state Theorem~\ref{thm:adaptive_regression} and Proposition~\ref{prop:query_refresh}, and we prove them together with Theorem~\ref{thm:adaptive_tree}. The two adversary models are defined in Section~\ref{sec:tree_data_structure}.

\begin{theorem}[Adaptive dynamic tensor product regression]\label{thm:adaptive_regression}
Work in exact real arithmetic with square roots, and run the private median of~\cite[Theorem~2.5]{bkm+22} in the time stated there. Let $b\in\R^n$ be the label vector, which is never updated, and let $\epsilon,\beta\in(0,1)$, $\xi>0$ and $T\in\mathbb{N}_+$. Fix one of the three cases of Theorem~\ref{thm:app:tree_approximation_guarantee}, that is, a pair $(C_{\rm base},T_{\rm base})$ together with its conditions on $m$ and $s_C$, and let $m$ and $s_C$ satisfy the conditions of that case with $d'=d+1$, with $\epsilon/72$ in place of $\epsilon$ and with failure probability $1/100$. There is a data structure {\sc AdaptiveDTRegression} (Algorithm~\ref{alg:adaptive_reg}) with the following guarantee against a standard adaptive adversary that makes any number of updates and at most $T$ queries. For the $t$-th query let $A^{(t)}$ be the current design matrix and $\opt_t:=\min_{x\in\R^d}\|A^{(t)}x-b\|_2$. With probability at least $1-\beta$, every answer $\hat x_t$ satisfies
\begin{align*}
\opt_t\le\|A^{(t)}\hat x_t-b\|_2\le\sqrt{1+\epsilon}\cdot\opt_t+\xi .
\end{align*}
In particular, $\|A^{(t)}\hat x_t-b\|_2\le(1+2\epsilon)\opt_t$ whenever $\opt_t\ge\xi/\epsilon$. If $\|b\|_2\le\xi$, the data structure answers $0$ and keeps no copies. Otherwise it keeps $K$ independent copies of {\sc DTRegression} (Algorithm~\ref{alg:kronecker_reg}, with a {\sc Query} that returns a minimizer chosen by a fixed rule) and the Gram matrices $A_i^\top A_i$, and it answers each query from $u$ sampled copies, where, with $L:=\log(4dT\|b\|_2/(\beta\xi))$,
\begin{align*}
u=O\big(\sqrt{d\log(4T/\beta)}\cdot L\big),\qquad K=O\big(\sqrt{Td}\cdot\log(4T/\beta)\cdot L\big).
\end{align*}
Let $t_b$ be the time to compute $\Pi^qb$ in Theorem~\ref{thm:app:dynamic_k_reg}. The running times are as follows, where $\wt O$ also hides factors polylogarithmic in $d$, $T$, $1/\beta$ and $\|b\|_2/\xi$.
\begin{itemize}
\item {\sc Initialize}: $K\cdot\big(\wt O(s_C\sum_{i=1}^q\nnz(A_i)+qmd)+t_b\big)+O\big(\nnz(b)+\sum_{i=1}^q(\nnz(A_i)d_i+d_i^3)\big)$.
\item {\sc Update}$(i,B)$: $K\cdot\wt O(s_C\nnz(B)+md)+O(\nnz(B)d_i+d_i^3)$.
\item {\sc Query}: $O(u\cdot(md^{\omega-1}+d^\omega))+\wt O(d^2)=\wt O(\sqrt d\cdot md^{\omega-1})$.
\end{itemize}
For {\sc CountSketch} leaves and {\sc TensorSketch} internal nodes we have $s_C=1$, $m=\Theta(\epsilon^{-2}qd^2)$ and $t_b=O(q\nnz(b))$. So {\sc Update} takes $\wt O(\sqrt{Td}\,(\nnz(B)+\epsilon^{-2}qd^3))+O(\nnz(B)d_i+d_i^3)$ time and {\sc Query} takes $\wt O(\epsilon^{-2}qd^{\omega+3/2})$ time.
\end{theorem}

\begin{proposition}[Query-time refresh]\label{prop:query_refresh}
Consider any of the data structures of Theorems~\ref{thm:app:dynamic_k_reg}, \ref{thm:app:spline} and~\ref{thm:app:low_rank}. Suppose that for every fixed input, a {\sc Query} made right after {\sc Initialize} meets its guarantee with probability at least $1-\delta_q$. Modify the structure as follows. {\sc Initialize} only stores the input. {\sc Update}$(i,B)$ only sets $A_i\gets A_i+B$. A {\sc Query} that is the first one or follows an update (a \emph{refreshing} query) runs {\sc Initialize} on the current input with fresh randomness and then {\sc Query}. Any other {\sc Query} returns the previous answer.
\begin{enumerate}
\item[(a)] Against a strong adaptive adversary that makes at most $T$ queries, all answers meet their guarantees with probability at least $1-T\delta_q$. {\sc Update} takes $\wt O(\nnz(B))$ time, and a refreshing {\sc Query} takes the time of {\sc Initialize} plus {\sc Query} of the original structure.
\item[(b)] For regression, let $\delta\in(0,1)$ and $\kappa:=\lceil8\ln(T/\delta)\rceil$. The structure also maintains $G_i=A_i^\top A_i$. A refreshing {\sc Query} builds $\kappa$ independent trees, each satisfying the {\sc Query} hypothesis of Theorem~\ref{thm:app:dynamic_k_reg} with $\epsilon/14$ in place of $\epsilon$ and with failure probability $1/4$, and computes their answers $x^{(1)},\ldots,x^{(\kappa)}$. It returns $x^{(\rho^*)}$, where $\rho^*$ minimizes over $\rho\in[\kappa]$ the $(\lfloor\kappa/2\rfloor+1)$-th smallest of the distances $\|A(x^{(\rho)}-x^{(\sigma)})\|_2$, $\sigma\in[\kappa]$, computed exactly from $G_1\otimes\cdots\otimes G_q$. Then, against a strong adaptive adversary that makes at most $T$ queries, with probability at least $1-\delta$ every answer satisfies $\|A\hat x-b\|_2\le(1+\epsilon)\opt$ for the current $A$. With {\sc CountSketch} leaves and {\sc TensorSketch} internal nodes this holds with $m=\Theta(\epsilon^{-2}qd^2)$, which depends on neither $T$ nor $\delta$. {\sc Update} takes $\wt O(\nnz(B)d_i)$ time. A refreshing {\sc Query} takes $\kappa\cdot\big(\wt O(s_C\sum_{i=1}^q\nnz(A_i)+qmd)+t_b+O(md^{\omega-1}+d^\omega)\big)+O(\kappa d\sum_{i=1}^qd_i+\kappa^2d)$ time.
\end{enumerate}
\end{proposition}

We use two general facts.

\begin{lemma}[Fresh randomness]\label{lem:fresh_randomness}
Let $\mathcal F$ be a $\sigma$-algebra, let $X$ be an $\mathcal F$-measurable random variable with values in a set $\mathcal I$, and let $R$ be a random variable independent of $\mathcal F$. Let $\mathcal B$ be a measurable set of pairs such that $\Pr_R[(x,R)\in\mathcal B]\le\delta_0$ for every fixed $x\in\mathcal I$. Then $\Pr[(X,R)\in\mathcal B\mid\mathcal F]\le\delta_0$ almost surely, and in particular $\Pr[(X,R)\in\mathcal B]\le\delta_0$.
\end{lemma}
\begin{proof}
Since $X$ is $\mathcal F$-measurable and $R$ is independent of $\mathcal F$, the freezing property of conditional expectation gives $\E[\mathbf 1_{\mathcal B}(X,R)\mid\mathcal F]=\phi(X)$ almost surely, where $\phi(x):=\Pr_R[(x,R)\in\mathcal B]\le\delta_0$.
\end{proof}

\begin{lemma}[Kronecker monotonicity]\label{lem:kron_monotone}
If $0\preceq P_1\preceq P_2$ and $0\preceq Q_1\preceq Q_2$, then $P_1\otimes Q_1\preceq P_2\otimes Q_2$.
\end{lemma}
\begin{proof}
We have $P_2\otimes Q_2-P_1\otimes Q_1=(P_2-P_1)\otimes Q_2+P_1\otimes(Q_2-Q_1)$, and the Kronecker product of two positive semidefinite matrices is positive semidefinite.
\end{proof}

\subsection{Proof of Theorem~\ref{thm:adaptive_tree}}

For a node $v=(k,\ell)$ let $M_v:=A_{k2^\ell+1}\otimes\cdots\otimes A_{(k+1)2^\ell}$ be the tensor product of the current factors below $v$, so that $M_{0,\log q}=A$. Let $D_v:=D_{k,\ell}$ be its number of columns, and let $n_v:=2^{\ell+1}-1$ be the number of nodes in the subtree of $v$.

\paragraph{Each draw sees a fixed input.}
The data structure draws $2q-1$ sketches in {\sc Initialize} and $1+\log q$ sketches in each update, so at most $N$ in total. Number the draws $1,2,\ldots$ in the order they are made. In {\sc Initialize} the leaves come first and then the internal nodes level by level, and in an update the nodes come from the leaf up. For $s\le N$ let $\mathcal F_s$ be the $\sigma$-algebra generated by the adversary's coins and the sketches of draws $1,\ldots,s-1$. Draw $s$ chooses the sketch $S_s$ of some node $v_s$ and sets $J_{v_s}\gets S_sX_s$. Here $X_s$ is the current $A_{k+1}$ if $v_s=(k,0)$ is a leaf, and $X_s=J_{2k,\ell-1}\otimes J_{2k+1,\ell-1}$, formed from the current children, if $v_s=(k,\ell)$ with $\ell\ge1$. Both $v_s$ and $X_s$ are $\mathcal F_s$-measurable, for three reasons. First, the operation in progress was chosen by the adversary from the answers and the random bits produced before it, and all of these are determined by the adversary's coins and the earlier draws. The same holds for the current factors. Second, in an update, the child of $v_s$ on the updated path was recomputed by an earlier draw of the same operation, and the other child was last computed by a draw of an earlier operation. Third, in {\sc Initialize}, both children were computed by earlier draws of {\sc Initialize}. The sketch $S_s$ is drawn fresh from the distribution of the node $v_s$. Formally, let $R_1,\ldots,R_N$ be independent of each other and of the adversary's coins, where $R_s$ consists of one independent sketch for every node, drawn from that node's distribution, and let $S_s$ be the entry of $R_s$ at $v_s$. This gives the sketches the same joint distribution as in the data structure. By induction on $s$, $\mathcal F_s$ is contained in the $\sigma$-algebra generated by the adversary's coins and $R_1,\ldots,R_{s-1}$. So $R_s$ is independent of $\mathcal F_s$, and Lemma~\ref{lem:fresh_randomness} applies with $X=(v_s,X_s)$ and $R=R_s$. The matrix $X_s$ has $n_{k+1}$ or $m^2$ rows and $D_{v_s}$ columns, so $\sd_0(X_s^\top X_s)=\rank(X_s)\le D_{v_s}$. Let $\mathcal E_s$ be the event that draw $s$ does not happen or
\begin{align}\label{eq:node_event}
(1-\epsilon')X_s^\top X_s\preceq(S_sX_s)^\top(S_sX_s)\preceq(1+\epsilon')X_s^\top X_s .
\end{align}
By the OSE hypothesis, for every fixed node and fixed input this fails with probability at most $\delta'$. By Lemma~\ref{lem:fresh_randomness}, $\Pr[\overline{\mathcal E_s}]\le\delta'$. A union bound over $s\le N$ shows that all events $\mathcal E_s$ hold with probability at least $1-N\delta'=1-\delta$. We assume this from now on. The old sketches off the updated path, which earlier answers may reveal, play no special role here: each of them only has to satisfy~\eqref{eq:node_event} for its own input.

\paragraph{Every current node satisfies its event.}
Fix a time between operations and a node $v$, and let $s$ be the draw that chose its current sketch. We claim that the current input of $v$ is still $X_s$, so that $J_v=S_sX_s$ satisfies~\eqref{eq:node_event}. The input of $v$ is a factor below $v$ or is formed from the matrices of its children. These change only in an update of a factor below $v$. Such an update lies on a path through $v$, so it first recomputes the child of $v$ on that path and then redraws $v$. Hence after draw $s$ the input of $v$ does not change until $v$ is redrawn.

\paragraph{Induction up the tree.}
We show that for every node $v$,
\begin{align}\label{eq:node_induction}
(1-\epsilon')^{n_v}M_v^\top M_v\preceq J_v^\top J_v\preceq(1+\epsilon')^{n_v}M_v^\top M_v .
\end{align}
For a leaf $v=(k,0)$, $n_v=1$ and $X_s=A_{k+1}=M_v$, so~\eqref{eq:node_induction} is~\eqref{eq:node_event}. Let $v$ be an internal node with children $L$ and $R$, for which~\eqref{eq:node_induction} holds. By the mixed-product property,
\begin{align*}
(J_L\otimes J_R)^\top(J_L\otimes J_R)= & ~ J_L^\top J_L\otimes J_R^\top J_R ,\\
M_L^\top M_L\otimes M_R^\top M_R= & ~ (M_L\otimes M_R)^\top(M_L\otimes M_R)=M_v^\top M_v .
\end{align*}
Lemma~\ref{lem:kron_monotone}, applied to the lower bounds and then to the upper bounds of~\eqref{eq:node_induction} at $L$ and $R$, gives
\begin{align*}
(1-\epsilon')^{n_L+n_R}M_v^\top M_v\preceq(J_L\otimes J_R)^\top(J_L\otimes J_R)\preceq(1+\epsilon')^{n_L+n_R}M_v^\top M_v .
\end{align*}
Combining this with~\eqref{eq:node_event} at $v$, where $X_s=J_L\otimes J_R$ and $J_v=S_sX_s$, and using $n_v=n_L+n_R+1$, gives~\eqref{eq:node_induction} at $v$. At the root, $n_v=2q-1$ and $(2q-1)\epsilon'=\epsilon/2$. So $(1+\epsilon')^{2q-1}\le e^{\epsilon/2}\le1+\epsilon$ for $\epsilon\in(0,1]$, and $(1-\epsilon')^{2q-1}\ge1-\epsilon/2$ by Bernoulli's inequality. This proves the guarantee.

\paragraph{The two instantiations.}
We use the following results of~\cite{akk+20}, numbered as in arXiv:\allowbreak1909.01410v5. A distribution over $S\in\R^{m\times D}$ has the $(\epsilon,\delta,t)$-JL moment property if $\E\|Sx\|_2^2=1$ and $\big(\E\,\big|\|Sx\|_2^2-1\big|^t\big)^{1/t}\le\epsilon\delta^{1/t}$ for every unit vector $x\in\R^D$.
\begin{itemize}
\item By Lemma~11 of~\cite{akk+20}, if $S$ has the $(\epsilon/\mu,\delta,t)$-JL moment property for some $t\ge1$, then $S$ is an $(\epsilon,\delta,\mu,D,n)$-OSE for every $n$.
\item By Lemma~18 and Remark~1 of~\cite{akk+20}, {\sc OSNAP} with $m=\Omega(\epsilon^{-2}\log(1/\delta))$ and sparsity $s=\Omega(\epsilon^{-1}\log(1/\delta))$ has the $(\epsilon,\delta,\log(1/\delta))$-JL moment property.
\item By Lemma~21 with $k=2$ and Remark~1 of~\cite{akk+20}, {\sc TensorSRHT} $\R^{m^2}\to\R^m$ with $m=\Omega(\epsilon^{-2}\log(1/\delta)\log^2(1/(\epsilon\delta)))$ has the same property. Lemma~21 holds for every unit vector of $\R^{m^2}$, not only for tensor products.
\item {\sc CountSketch} and degree-two {\sc TensorSketch} with $m=\Omega(\epsilon^{-2}/\delta)$ have the $(\epsilon,\delta,2)$-JL moment property for every unit vector, as used in the proof of Theorem~1 of~\cite{akk+20}.
\end{itemize}
We apply these at each node $v$ with $\epsilon'/D_v$ in place of $\epsilon$, $\delta'$ in place of $\delta$ and $\mu=D_v$. Here $D_v=d_i$ at the leaf of $A_i$, and $D_v\le d$ at every node. Since $1/\epsilon'=2(2q-1)/\epsilon\le4q/\epsilon$ and $1/\delta'=N/\delta$, this gives the stated conditions on $m$ and $s_{C,i}$.

\paragraph{Running time.}
{\sc Update}$(i,B)$ stores $A_i+B$ in $\wt O(\nnz(B))$ time. It draws the leaf sketch, generating only the columns that meet nonzero rows of $A_i+B$, and computes the new leaf in $O(s_{C,i}(\nnz(A_i)+\nnz(B))+md_i)$ time, as in the proof of Theorem~\ref{thm:app:tree_init}. At a node $v$ on the path with children $L$ and $R$, each of the $D_v$ columns of $J_L\otimes J_R$ is the tensor product of a column of $J_L$ and a column of $J_R$. So applying a fresh {\sc TensorSketch} or {\sc TensorSRHT} costs $O(D_vm\log m)$, as in the proof of Theorem~\ref{thm:app:tree_init}. There are $\log q$ such nodes and $D_v\le d$, so the path costs $O(md\log m\log q)=\wt O(md)$. {\sc Initialize} is that of Theorem~\ref{thm:app:tree_init}, with sparsity $s_{C,i}$ at the leaf of $A_i$, and it also stores the factors. \qed

\subsection{Proof of Theorem~\ref{thm:adaptive_regression}}

We cite the results of~\cite{bkm+22} by their numbers in arXiv:\allowbreak2111.03980v1.

\paragraph{Parameters.}
Set $\delta_{\rm DP}:=\beta/(400T)$, $\delta_1:=\delta_{\rm DP}/(2T)$ and $\epsilon_{\rm DP}:=1/(2\sqrt{2d\ln(1/\delta_1)})$. Let $N_{\mathcal X}:=\lceil2\sqrt{2d}\,\|b\|_2/\xi\rceil$ and $h:=2\|b\|_2/N_{\mathcal X}\le\xi/\sqrt{2d}$, and let $\mathcal X:=\{-\|b\|_2+jh: j=0,1,\ldots,N_{\mathcal X}\}$, a grid of $[-\|b\|_2,\|b\|_2]$ with step $h$. Let $\beta':=\beta/(4dT)$. \textsc{PMedian}$_{\epsilon_{\rm DP},\beta'}$ denotes the private median of~\cite[Theorem~2.5]{bkm+22} over the domain $\mathcal X$. It is $(\epsilon_{\rm DP},0)$-differentially private. On any input of $u$ points of $\mathcal X$, with probability at least $1-\beta'$, it outputs a point $g\in\mathcal X$ such that at least $u/2-\Gamma$ of the inputs are $\ge g$ and at least $u/2-\Gamma$ of them are $\le g$, where $\Gamma=O(\epsilon_{\rm DP}^{-1}\log(|\mathcal X|/\beta'))$. Set
\begin{align}\label{eq:adaptive_params}
u:=\big\lceil\max\{20\Gamma,\ 62\ln(4T/\beta)\}\big\rceil ,\qquad K:=\big\lceil1200\,u\sqrt{2T\ln(2/\delta_{\rm DP})}\big\rceil .
\end{align}
If $\|b\|_2>\xi$, then $|\mathcal X|\le4\sqrt{2d}\,\|b\|_2/\xi$ and $\epsilon_{\rm DP}^{-1}=O(\sqrt{d\log(4T/\beta)})$. Since also $L\ge\ln(4T/\beta)>1$, this gives the bounds on $u$ and $K$ in the theorem.

\begin{algorithm}[!ht]\caption{Adaptive dynamic tensor product regression}\label{alg:adaptive_reg}
\begin{algorithmic}[1]
\State {\bf data structure} {\sc AdaptiveDTRegression} \Comment{Theorem~\ref{thm:adaptive_regression}}
\State
\State {\bf members}
\State \hspace{4mm} {\sc DTRegression} $\mathcal D_1,\ldots,\mathcal D_K$ \Comment{Independent copies of Algorithm~\ref{alg:kronecker_reg}}
\State \hspace{4mm} $A_i$, $G_i\in\R^{d_i\times d_i}$, $F_i\in\R^{r_i\times d_i}$ and $F_i^+$ for $i\in[q]$, and $\|b\|_2$ \Comment{$G_i=A_i^\top A_i=F_i^\top F_i$}
\State {\bf end members}
\State
\Procedure{Initialize}{$A_1,\ldots,A_q,m,C_{\rm base},T_{\rm base},b$}
\State Compute $\|b\|_2$, and if $\|b\|_2\le\xi$, set $K\gets0$ \Comment{Then no copies are kept}
\For{$j=1\to K$}
\State $\mathcal D_j.\textsc{Initialize}(A_1,\ldots,A_q,m,C_{\rm base},T_{\rm base},b)$ \Comment{With fresh independent randomness $\rho_j$}
\EndFor
\For{$i=1\to q$}
\State $G_i\gets A_i^\top A_i$, and compute $F_i$ and $F_i^+$
\EndFor
\EndProcedure
\State
\Procedure{Update}{$i\in[q],B\in\R^{n_i\times d_i}$}
\For{$j=1\to K$}
\State $\mathcal D_j.\textsc{Update}(i,B)$
\EndFor
\State $G_i\gets G_i+A_i^\top B+B^\top A_i+B^\top B$, then $A_i\gets A_i+B$, and recompute $F_i$ and $F_i^+$
\EndProcedure
\State
\Procedure{Query}{$ $}
\If{$\|b\|_2\le\xi$ or $r_i=0$ for some $i\in[q]$}
\State \Return $0$
\EndIf
\State Draw $j_1,\ldots,j_u$ independently and uniformly from $[K]$
\For{$p=1\to u$}
\State $x_p\gets\mathcal D_{j_p}.\textsc{Query}()$, and $y_p\gets\mathrm{rd}\big((F_1\otimes\cdots\otimes F_q)x_p\big)$
\EndFor
\For{$l=1\to r$}
\State $g_l\gets\textsc{PMedian}_{\epsilon_{\rm DP},\beta'}(y_{1,l},\ldots,y_{u,l})$ \Comment{Private median over the domain $\mathcal X$}
\EndFor
\State \Return $(F_1^+\otimes\cdots\otimes F_q^+)\,g$
\EndProcedure
\end{algorithmic}
\end{algorithm}

\paragraph{Notation and the copies.}
Throughout, $A=A_1\otimes\cdots\otimes A_q$ is the design matrix at the query under consideration, $\opt:=\min_x\|Ax-b\|_2$, $r_i:=\rank(A_i)$ and $r:=\prod_{i=1}^qr_i\le d$. The factor $F_i\in\R^{r_i\times d_i}$ has full row rank and satisfies $F_i^\top F_i=G_i$. In exact arithmetic with square roots, Cholesky factorization with symmetric pivoting computes it in $O(d_i^3)$ time~\cite{golubVanLoan2013Book}, and $F_i^+=F_i^\top(F_iF_i^\top)^{-1}$. Let $W:=F_1\otimes\cdots\otimes F_q\in\R^{r\times d}$. The map $\mathrm{rd}$ acts coordinatewise. It clips a number to $[-\|b\|_2,\|b\|_2]$ and rounds it to the nearest point of $\mathcal X$, breaking ties toward the smaller point. Using comparisons only, this is a binary search over the indices $j$ in $O(\log|\mathcal X|)$ time, and $N_{\mathcal X}$ is found the same way. Each copy is {\sc DTRegression} with one change. Its {\sc Query} returns the minimizer of $\|Mx-\wt b\|_2$ supported on the first maximal set $S$ of linearly independent columns of $M$, that is, $x_S=(M_S^\top M_S)^{-1}M_S^\top\wt b$ and $x_{[d]\setminus S}=0$. This is a minimizer because the columns of $M_S$ span the column space of $M$. The columns of $M^\top M$ satisfy the same linear relations as those of $M$, because $M^\top My=0$ exactly when $My=0$. Since $M^\top M$ is symmetric, $S$ is also the first maximal set of linearly independent rows of $M^\top M$, which the LSP decomposition of~\cite{imh82} finds in $O(d^\omega)$ time. So the {\sc Query} time stays $O(md^{\omega-1}+d^\omega)$. The change makes each answer a fixed function of $(M,\wt b)$, even when $M^\top M$ is singular.

\paragraph{Trivial cases.}
If $\|b\|_2\le\xi$, the answer $0$ satisfies $\|A\cdot0-b\|_2=\|b\|_2\le\xi$. If $r_i=0$ for some $i$, then $A=0$ and every $x$ is optimal. In both cases the answer is a function of the earlier answers and the adversary's choices alone. From now on we assume $\|b\|_2>\xi$ and $r_i\ge1$ for all $i$.

\begin{lemma}[Whitening]\label{lem:whitening}
\begin{enumerate}
\item[(i)] $W^\top W=A^\top A$.
\item[(ii)] $W$ has full row rank, so $WW^+=I_r$, and $W^+=F_1^+\otimes\cdots\otimes F_q^+$.
\item[(iii)] The vector $z^*:=Wx^*$ is the same for every minimizer $x^*$ of $\|Ax-b\|_2$, and $|z^*_l|\le\|z^*\|_2\le\|b\|_2$ for all $l\in[r]$.
\item[(iv)] For every $x\in\R^d$, $\|Ax-b\|_2^2=\opt^2+\|Wx-z^*\|_2^2$.
\end{enumerate}
\end{lemma}
\begin{proof}
(i) By the mixed-product property, $W^\top W=\bigotimes_{i=1}^qF_i^\top F_i=\bigotimes_{i=1}^qA_i^\top A_i=A^\top A$. (ii) By the mixed-product property, $WW^\top=\bigotimes_{i=1}^qF_iF_i^\top$. Each $F_iF_i^\top$ is invertible because $F_i$ has full row rank, and again by the mixed-product property $\bigotimes_{i=1}^q(F_iF_i^\top)^{-1}$ is the inverse of $WW^\top$. So the $r\times r$ matrix $WW^\top$ is invertible, and $W$ has full row rank $r$. Hence $W^+=W^\top(WW^\top)^{-1}=\bigotimes_{i=1}^qF_i^\top(F_iF_i^\top)^{-1}=\bigotimes_{i=1}^qF_i^+$. (iii) By (i), $\|Wy\|_2=\|Ay\|_2$ for every $y$, so $Wy=0$ exactly when $Ay=0$. Two minimizers differ by a vector $y$ with $Ay=0$, so they have the same image under $W$. The normal equations give $(Ax^*)^\top(Ax^*-b)=0$. Hence $\|b\|_2^2=\|Ax^*\|_2^2+\opt^2$ and $\|z^*\|_2=\|Ax^*\|_2\le\|b\|_2$. (iv) Expanding around $x^*$ and using $A^\top(Ax^*-b)=0$ gives $\|Ax-b\|_2^2=\opt^2+\|A(x-x^*)\|_2^2$, which equals $\opt^2+\|Wx-z^*\|_2^2$ by (i).
\end{proof}

\begin{lemma}[Coordinatewise median in a ball]\label{lem:median_ball}
Let $y_1,\ldots,y_u,c\in\R^r$, and let $\mathcal G\subseteq[u]$ satisfy $|\mathcal G|\ge(1-a)u$ and $\|y_p-c\|_2\le\varrho$ for all $p\in\mathcal G$. Let $g\in\R^r$ be such that, for every $l\in[r]$, at least $\gamma u$ of the numbers $y_{1,l},\ldots,y_{u,l}$ are $\ge g_l$ and at least $\gamma u$ of them are $\le g_l$, where $\gamma>a$. Then $\|g-c\|_2\le\varrho/\sqrt{\gamma-a}$.
\end{lemma}
\begin{proof}
Fix $l$ with $g_l>c_l$, and let $P_l:=\{p\in\mathcal G: y_{p,l}\ge g_l\}$. Then $|P_l|\ge\gamma u-(u-|\mathcal G|)\ge(\gamma-a)u$, and every $p\in P_l$ has $y_{p,l}-c_l\ge g_l-c_l>0$. Averaging over $P_l$ gives
\begin{align*}
(g_l-c_l)^2\le\frac{1}{|P_l|}\sum_{p\in P_l}(y_{p,l}-c_l)^2\le\frac{1}{(\gamma-a)u}\sum_{p\in\mathcal G}(y_{p,l}-c_l)^2 .
\end{align*}
The case $g_l<c_l$ is symmetric, and the bound is trivial when $g_l=c_l$. Summing over $l\in[r]$ gives $\|g-c\|_2^2\le\frac{1}{(\gamma-a)u}\sum_{p\in\mathcal G}\|y_p-c\|_2^2\le\frac{\varrho^2}{\gamma-a}$.
\end{proof}
The bound does not depend on the dimension $r$. This is why we take medians in the coordinates $z=Wx$, where by Lemma~\ref{lem:whitening}(iv) $\|Ax-b\|_2^2-\opt^2$ is the squared Euclidean distance from $z$ to $z^*$.

\begin{lemma}[One copy on a fixed input]\label{lem:one_copy}
For a copy with random string $\rho$, let $x(\rho)$ be its answer to a query. Then $x(\rho)$ depends only on the current factors, $b$ and $\rho$. For every fixed $A_1,\ldots,A_q$ and $b$, with probability at least $99/100$ over $\rho$,
\begin{align*}
\|Ax(\rho)-b\|_2\le(1+\epsilon/24)\,\opt\qquad\text{and}\qquad\|Wx(\rho)-z^*\|_2^2\le(\epsilon/8)\,\opt^2 .
\end{align*}
\end{lemma}
\begin{proof}
Let $\Pi_\rho$ be the sketching matrix formed by the sketches of the copy. As shown in the proof of Theorem~\ref{thm:app:dynamic_k_reg}, after any sequence of updates the root of the copy stores $M=\Pi_\rho A$, and $\wt b=\Pi_\rho b$. The answer is a fixed function of $(M,\wt b)$, which proves the first claim. Apply Theorem~\ref{thm:app:tree_approximation_guarantee} to the fixed matrix $[A,~b]$ with $d'=d+1$, accuracy $a:=\epsilon/72$ and failure probability $1/100$. With probability at least $99/100$ we have $\|\Pi_\rho(Ax-b)\|_2=(1\pm a)\|Ax-b\|_2$ for all $x$. On this event, as in the proof of Theorem~\ref{thm:app:dynamic_k_reg},
\begin{align*}
\|Ax(\rho)-b\|_2\le & ~ \frac{\|\Pi_\rho(Ax(\rho)-b)\|_2}{1-a}\le\frac{\|\Pi_\rho(Ax^*-b)\|_2}{1-a}\\
\le & ~ \frac{1+a}{1-a}\,\opt\le(1+3a)\,\opt=(1+\epsilon/24)\,\opt .
\end{align*}
By Lemma~\ref{lem:whitening}(iv), $\|Wx(\rho)-z^*\|_2^2\le((1+\epsilon/24)^2-1)\opt^2=(\epsilon/12+\epsilon^2/576)\opt^2\le(\epsilon/8)\opt^2$.
\end{proof}

\begin{lemma}[Privacy of the random strings]\label{lem:adaptive_privacy}
Let $R=(\rho_1,\ldots,\rho_K)$ be the random strings of the copies, and call two such vectors neighbors if they differ in one entry. For every adversary and every fixing of its coins, the sequence of answers is $(1/100,\delta_{\rm DP})$-differentially private as a function of $R$.
\end{lemma}
\begin{proof}
Updates produce no output. The adversary's choices are functions of earlier answers and of its coins, which are independent of $R$. So the sequence of answers is an adaptive composition of at most $T$ mechanisms, one per query, where the mechanism of a query is determined by the earlier answers. Fix a query and its history. This fixes $A$ and $W$, and also whether the answer is the trivial one, which does not depend on $R$. Otherwise the query samples $u$ entries of $R$ with replacement and runs a mechanism $\mathcal M$ on them. The map $\rho\mapsto\mathrm{rd}(Wx(\rho))$ acts on each sampled string separately. So changing one sampled string changes one input of each of the $r\le d$ calls to \textsc{PMedian}. Each call is $(\epsilon_{\rm DP},0)$-differentially private. By advanced composition~\cite{drv10} (as stated in~\cite[Theorem~2.2]{bkm+22}) with $k=d$ and $\delta'=\delta_1$, the mechanism $\mathcal M$ is $(\epsilon_c,\delta_1)$-differentially private with
\begin{align*}
\epsilon_c=\sqrt{2d\ln(1/\delta_1)}\,\epsilon_{\rm DP}+2d\epsilon_{\rm DP}^2=\frac12+\frac{1}{4\ln(1/\delta_1)}\le\frac34 .
\end{align*}
(If $r<d$, we compose fewer mechanisms, which only helps.) Since $\epsilon_c\le1$ and $K\ge2u$, the subsampling lemma of~\cite{bnsv15} (Lemma~4.14 of arXiv:\allowbreak1504.07553v2, see also~\cite[Theorem~2.3]{bkm+22}) shows that the query is $(\epsilon_r,\delta_r)$-differentially private as a function of $R$, with
\begin{align*}
\epsilon_r=\frac{6\epsilon_cu}{K}\le\frac{4.5\,u}{K}\le\frac{4.5}{1200\sqrt{2T\ln(2/\delta_{\rm DP})}}\le0.004 ,\qquad\delta_r=e^{\epsilon_r}\cdot\frac{4u}{K}\cdot\delta_1\le\delta_1 .
\end{align*}
Advanced composition over at most $T$ queries with $\delta'=\delta_{\rm DP}/2$ gives privacy parameters
\begin{align*}
\sqrt{2T\ln(2/\delta_{\rm DP})}\,\epsilon_r+2T\epsilon_r^2\le\frac{4.5}{1200}+\frac{4.5^2}{1200^2\ln(2/\delta_{\rm DP})}<\frac{1}{100}\qquad\text{and}\qquad\frac{\delta_{\rm DP}}{2}+T\delta_1=\delta_{\rm DP} .\qedhere
\end{align*}
\end{proof}

\paragraph{Accuracy.}
For the $t$-th query let $f_t(\rho):=1$ if $\|A^{(t)}x(\rho)-b\|_2\le(1+\epsilon/24)\opt_t$, and $f_t(\rho):=0$ otherwise. We show that (E1) fails with probability at most $\beta/4$, that (E1) holds while (E2) fails with probability at most $\beta/4$, and that (E3) fails with probability at most $\beta/4$.
\begin{itemize}
\item[(E1)] For every query $t$, $\frac1K\sum_{j=1}^Kf_t(\rho_j)\ge0.89$.
\item[(E2)] For every query $t$, at least $0.8u$ of the sampled indices $j_1,\ldots,j_u$ have $f_t(\rho_{j_p})=1$.
\item[(E3)] Every call to \textsc{PMedian} meets its guarantee.
\end{itemize}

\emph{(E1).} Fix the adversary's coins and a query $t$. The design matrix $A^{(t)}$, and hence the predicate $f_t$, is a function of the answers to the earlier queries. By Lemma~\ref{lem:adaptive_privacy} and post-processing, the map $R\mapsto f_t$ is $(1/100,\delta_{\rm DP})$-differentially private. The strings $\rho_j$ are independent and identically distributed. By Lemma~\ref{lem:one_copy}, $\E_\rho f_t(\rho)\ge0.99$ for every possible value of $A^{(t)}$. We apply the generalization theorem of differential privacy~\cite{dfh+15,bns+21} (as stated in~\cite[Theorem~2.4]{bkm+22}) with privacy parameters $(1/100,\delta_{\rm DP})$, where $\delta_{\rm DP}=\beta/(400T)<1/400$. Its sample-size condition $K\ge10^4\log(8T/\beta)$ holds, because $K\ge1200u\ge74400\ln(4T/\beta)$, while $10^4\log_2(8T/\beta)\le\frac{10^4}{\ln2}\cdot\frac32\ln(4T/\beta)<21700\ln(4T/\beta)$ by $\ln2\le\frac12\ln(4T/\beta)$. So $\frac1K\sum_jf_t(\rho_j)<0.99-0.1$ with probability less than $100\,\delta_{\rm DP}=\beta/(4T)$. A union bound over at most $T$ queries bounds the failure probability of (E1) by $\beta/4$.

\emph{(E2).} The indices $j_1,\ldots,j_u$ of query $t$ are drawn after $A^{(t)}$ is fixed, independently of everything else. So, given $R$ and the history, the values $f_t(\rho_{j_p})$ are independent Bernoulli variables with mean $\frac1K\sum_jf_t(\rho_j)$, which is at least $0.89$ if (E1) holds at query $t$. By Hoeffding's inequality, fewer than $0.8u$ of them equal $1$ with probability at most $\exp(-2u\cdot0.09^2)\le\exp(-0.0162\cdot62\ln(4T/\beta))\le\beta/(4T)$. A union bound over the queries gives $\beta/4$.

\emph{(E3).} Each call fails with probability at most $\beta'=\beta/(4dT)$ whatever its input, and there are at most $dT$ calls.

Assume (E1), (E2) and (E3), and fix a nontrivial query $t$, with $A=A^{(t)}$ and $\opt=\opt_t$. Let $\mathcal G:=\{p\in[u]: f_t(\rho_{j_p})=1\}$, so $|\mathcal G|\ge0.8u$. For $p\in\mathcal G$, we have $x_p=x(\rho_{j_p})$ and $f_t(\rho_{j_p})=1$, so Lemma~\ref{lem:whitening}(iv), applied as in the last step of the proof of Lemma~\ref{lem:one_copy}, gives $\|Wx_p-z^*\|_2\le\sqrt{\epsilon/8}\,\opt$. By Lemma~\ref{lem:whitening}(iii), $z^*_l$ lies in $[-\|b\|_2,\|b\|_2]$, so clipping a coordinate to this interval does not increase its distance to $z^*_l$. Rounding to $\mathcal X$ moves a point of the interval by at most $h/2$. So $|y_{p,l}-z^*_l|\le|(Wx_p)_l-z^*_l|+h/2$ for every $l$, and the triangle inequality gives
\begin{align*}
\|y_p-z^*\|_2\le\sqrt{\epsilon/8}\,\opt+\sqrt r\,h/2=:\varrho\qquad\text{for all }p\in\mathcal G .
\end{align*}
By (E3) and $\Gamma\le u/20$, for each coordinate $l$ at least $(1/2-1/20)u$ of the values $y_{p,l}$ lie on each side of $g_l$. Lemma~\ref{lem:median_ball} with $a=1/5$ and $\gamma=9/20$ gives $\|g-z^*\|_2\le2\varrho=\sqrt{\epsilon/2}\,\opt+\sqrt r\,h$. The answer is $\hat x=W^+g$, so $W\hat x=WW^+g=g$ by Lemma~\ref{lem:whitening}(ii). By Lemma~\ref{lem:whitening}(iv), $(x+y)^2\le2x^2+2y^2$, $r\le d$ and $h\le\xi/\sqrt{2d}$,
\begin{align*}
\|A\hat x-b\|_2^2=\opt^2+\|g-z^*\|_2^2\le\opt^2+\epsilon\,\opt^2+2rh^2\le(1+\epsilon)\opt^2+\xi^2 .
\end{align*}
Hence $\|A\hat x-b\|_2\le\sqrt{1+\epsilon}\,\opt+\xi$. The lower bound $\opt\le\|A\hat x-b\|_2$ holds by definition. If $\opt\ge\xi/\epsilon$, then $\sqrt{1+\epsilon}\,\opt+\xi\le(1+\epsilon/2)\opt+\epsilon\,\opt\le(1+2\epsilon)\opt$. All three events hold with probability at least $1-3\beta/4\ge1-\beta$.

\paragraph{Running time.}
{\sc Initialize} runs $K$ copies, each in $\wt O(s_C\sum_i\nnz(A_i)+qmd)+t_b$ time by Theorem~\ref{thm:app:dynamic_k_reg}. Computing $G_i=A_i^\top A_i$ takes $O(\nnz(A_i)d_i)$ time, because a row with $k$ nonzero entries contributes $k^2\le kd_i$ products. Computing $F_i$ and $F_i^+$ takes $O(d_i^3)$ time, and computing $\|b\|_2$ takes $O(\nnz(b))$ time. {\sc Update} runs $K$ copy updates. The products $A_i^\top B$ and $B^\top B$ take $O(\nnz(B)d_i)$ time, since each nonzero entry of $B$ meets one row of $A_i$ or of $B$, of length $d_i$, and refactoring takes $O(d_i^3)$ time. {\sc Query} runs $u$ copy queries in $O(u(md^{\omega-1}+d^\omega))$ time. Each product $(F_1\otimes\cdots\otimes F_q)x_p$, and the final product with $F_1^+\otimes\cdots\otimes F_q^+$, takes $O(d\sum_id_i)$ time by $q$ mode products. This is $O(d^2+dq)$ since $\sum_id_i\le d+q-1$, and hence it is within $O(md^{\omega-1}+d^\omega)$ since $q\le m$. The roundings take $O(ur\log|\mathcal X|)=\wt O(d^{3/2})$ time. The $r\le d$ calls to \textsc{PMedian} take $O(u\,\epsilon_{\rm DP}^{-1}\log^3(|\mathcal X|/\beta')\,\mathrm{polylog}(u))=\wt O(u\sqrt d)$ time each, so $\wt O(ud^{3/2})=\wt O(d^2)$ in total. Finally, $u=\wt O(\sqrt d)$, and $d^\omega\le md^{\omega-1}$ since $m\ge d$. For {\sc CountSketch} and {\sc TensorSketch}, Theorem~\ref{thm:app:tree_approximation_guarantee} with $d'=d+1$, accuracy $\epsilon/72$ and failure probability $1/100$ needs $m=\Theta(\epsilon^{-2}qd^2)$, and $t_b=O(q\nnz(b))$ by Theorem~\ref{thm:app:dynamic_k_reg}. This gives the last claim. \qed

\subsection{Proof of Proposition~\ref{prop:query_refresh}}

\begin{lemma}[Selection]\label{lem:selection}
Let $x^*$ minimize $\|Ax-b\|_2$, and let $Z:=\|Ax^*-b\|_2$. Let $x^{(1)},\ldots,x^{(\kappa)}\in\R^d$, and let $\mathcal G\subseteq[\kappa]$ with $|\mathcal G|>\kappa/2$ and $\|A(x^{(\rho)}-x^*)\|_2\le\eta$ for all $\rho\in\mathcal G$. Let $\Delta_\rho$ be the $(\lfloor\kappa/2\rfloor+1)$-th smallest of the numbers $\|A(x^{(\rho)}-x^{(\sigma)})\|_2$, $\sigma\in[\kappa]$, and let $\rho^*$ minimize $\Delta_\rho$. Then $\|Ax^{(\rho^*)}-b\|_2^2\le Z^2+9\eta^2$.
\end{lemma}
\begin{proof}
For $\rho\in\mathcal G$, every $\sigma\in\mathcal G$ has $\|A(x^{(\rho)}-x^{(\sigma)})\|_2\le2\eta$, and $|\mathcal G|\ge\lfloor\kappa/2\rfloor+1$. So $\Delta_\rho\le2\eta$, and hence $\Delta_{\rho^*}\le2\eta$. The set $\{\sigma:\|A(x^{(\rho^*)}-x^{(\sigma)})\|_2\le\Delta_{\rho^*}\}$ has at least $\lfloor\kappa/2\rfloor+1>\kappa/2$ elements, so it meets $\mathcal G$ in some $\sigma$. Then $\|A(x^{(\rho^*)}-x^*)\|_2\le\Delta_{\rho^*}+\eta\le3\eta$. Since $b-Ax^*$ is orthogonal to the column space of $A$, we have $\|Ax-b\|_2^2=Z^2+\|A(x-x^*)\|_2^2$ for every $x$.
\end{proof}

\begin{proof}[Proof of Proposition~\ref{prop:query_refresh}]
Number the refreshing queries $1,2,\ldots$. There are at most $T$ of them. For the $k$-th one, let $\mathcal F_k$ be the $\sigma$-algebra generated by the adversary's coins and all random bits drawn before it. The current input is $\mathcal F_k$-measurable, because the adversary chose the updates from information in $\mathcal F_k$, even if it sees all random bits drawn so far. The sketches drawn by the $k$-th refreshing query are independent of $\mathcal F_k$. A query that does not refresh repeats the previous answer on the same input, so it meets its guarantee exactly when that answer does.

(a) By the hypothesis and Lemma~\ref{lem:fresh_randomness}, the $k$-th refreshing query fails with probability at most $\delta_q$. A union bound over at most $T$ refreshing queries gives $T\delta_q$.

(b) Let $\epsilon_1:=\epsilon/14$, and call a tree good if its answer $x$ satisfies $\|Ax-b\|_2\le(1+\epsilon_1)\opt$. By Theorem~\ref{thm:app:dynamic_k_reg}, applied with $\epsilon_1$ in place of $\epsilon$ and failure probability $1/4$, each tree is good with probability at least $3/4$ for every fixed input. Given $\mathcal F_k$, the input is fixed and the $\kappa$ trees are independent of $\mathcal F_k$ and of each other. So each is good with probability at least $3/4$, independently. By Hoeffding's inequality, at most $\kappa/2$ trees are good with probability at most $\exp(-2\kappa/16)=\exp(-\kappa/8)\le\delta/T$. Otherwise Lemma~\ref{lem:selection} applies with $\mathcal G$ the set of good trees and $\eta^2=3\epsilon_1\opt^2$, because a good tree has $\|A(x-x^*)\|_2^2=\|Ax-b\|_2^2-\opt^2\le((1+\epsilon_1)^2-1)\opt^2\le3\epsilon_1\opt^2$. It gives $\|A\hat x-b\|_2^2\le(1+27\epsilon_1)\opt^2\le(1+2\epsilon)\opt^2\le(1+\epsilon)^2\opt^2$. A union bound over at most $T$ refreshing queries gives $\delta$. The tree hypothesis is that of Theorem~\ref{thm:app:tree_approximation_guarantee} with $d'=d+1$, accuracy $\epsilon_1/3=\epsilon/42$ and failure probability $1/4$. For {\sc CountSketch} and {\sc TensorSketch} it holds with $m=\Theta(\epsilon^{-2}qd^2)$.

Running time. Each tree costs one {\sc Initialize} of Theorem~\ref{thm:app:dynamic_k_reg} and one {\sc Query}. For the selection, $\|A(x-x')\|_2^2=(x-x')^\top(G_1\otimes\cdots\otimes G_q)(x-x')$. Computing $(G_1\otimes\cdots\otimes G_q)x^{(\rho)}$ by $q$ mode products takes $O(d\sum_id_i)$ time per candidate, and then each of the $\kappa^2$ distances takes $O(d)$ time. Updating $G_i$ takes $O(\nnz(B)d_i)$ time, as in the proof of Theorem~\ref{thm:adaptive_regression}, and storing $A_i+B$ takes $\wt O(\nnz(B))$ time.
\end{proof}

\section{Omitted details and proofs for Dynamic Tensor Spline Regression (section~\ref{sec:spline_regression})}\label{sec:appendix-splines}

To specify the sketching dimension $m$ needed , we use the following definition of {\it Statistical Dimension} for Splines~\cite{dssw18}:

\begin{definition}[Statistical Dimension for Splines]
For the Spline regression problem specified by Eq.~\ref{eq:spline_regression}, the statistical dimension is defined as $\sd_{\lambda}(A,L) = \sum_{i}1/(1 + \lambda/\gamma_i^2) + d - p$. Here $\{\gamma_i, i \in [p]\}$ are the {\it generalized singular values} of $(A,L)$ which are defined as follows.
\end{definition}

\begin{definition}[Generalized singular values]
For matrices $A \in \R^{n \times d}$ and $L \in \R^{p \times d}$ such that $\rank(L) = p$ and $\rank\left(\mat{c} A \\ L \rix\right) = d$, the generalized singular value decomposition (GSVD)~\cite{golubVanLoan2013Book} of $(A,L)$ is given by the pair of factorizations $A=U\begin{bmatrix}\Sigma& {0}_{p\times (d-p)}\cr { 0}_{(d-p)\times p}& I_{d-p}
\end{bmatrix}RQ^\top  \quad \hbox{and} \quad
L=V\begin{bmatrix}\Omega  &{
0}_{p \times (d-p)}\end{bmatrix}RQ^\top, 
$
where $U \in \R^{n\times d}$ has orthonormal columns,
$V \in \R^{p \times p}$, $Q\in \R^{d \times d} $ are orthogonal, $R\in\mathbb{R}^{d\times d}$ is
upper triangular and nonsingular, and $\Sigma$ and $\Omega$ are
$p\times p$ diagonal matrices:
$\Sigma=\diag(\sigma_1,\sigma_2,\ldots,\sigma_{p})$ and
$\Omega=\diag(\mu_1,\mu_2,\ldots,\mu_p)$ with
$
0 \leq \sigma_1 \leq \sigma_2 \leq \ldots \leq \sigma_p < 1
\quad \hbox{and} \quad
 1 \geq \mu_1 \geq \mu_2 \geq \ldots \geq \mu_p>0,
$
satisfying $\Sigma^\top \Sigma+\Omega^\top \Omega=I_{p}$.  The {\em generalized singular values $\gamma_i$} of $(A,L)$ are defined by the ratios $\gamma_i=\sigma_i/\mu_i$ for $i \in [p]$. 

\end{definition}

We prove that for the spline regression, our sketching matrix can have even smaller dimension. The following is due to~\cite{dssw18}.

\begin{lemma}
\label{lem:dssw18_1}
Let $x^*=\arg\min_{x\in \R^d} \|Ax-b\|_2^2+\lambda \|Lx\|_2^2$, $A\in \R^{n\times d}$ and  $b\in \R^n$. Let $U_1\in \R^{n\times d}$ be the first $n$ rows of an orthogonal basis for $\begin{bmatrix}
A \\
\sqrt{\lambda} L
\end{bmatrix}\in \R^{(n+p)\times d}$. Let sketching matrix $S\in \R^{m\times n}$ have a distribution such that with probability $1-\delta$,
\begin{itemize}
    \item $\|U_1^\top S^\top SU_1-U_1^\top U_1\|_2 \leq 1/4$,
    \item $\|U_1^\top(S^\top S-I)(b-Ax^*)\|_2 \leq \sqrt{\epsilon~\mathrm{OPT}/2}$.
\end{itemize}
Let $\wt x$ denote $\arg\min_{x\in \R^d} \|S(Ax-b)\|_2^2+\lambda \|Lx\|_2^2$. Then with probability at least $1-\delta$,
\begin{align*}
    \|A\wt x-b\|_2^2 + \lambda \|L\wt x\|_2^2 = & ~ (1\pm\epsilon)\mathrm{OPT}.
\end{align*}
\end{lemma}

\begin{lemma}
\label{lem:dssw18_2}
For $U_1$ as in Lemma~\ref{lem:dssw18_1}, we have 
\begin{align*}
    \|U_1\|_F^2 = & ~ \sd_{\lambda}(A, L).
\end{align*}
\end{lemma}

\begin{algorithm}[!ht]\caption{Dynamic Tensor Spline Regression}
\label{alg:kronecker_spline}
\begin{algorithmic}[1]
\State {\bf data structure} {\sc DTSRegression}
\State
\State {\bf members}
\State \hspace{4mm} $\textsc{DynamicTensorTree DTT} $
\State \hspace{4mm} $A_1\in \R^{n_1\times d_1},\ldots,A_q\in \R^{n_q\times d_q}$
\State \hspace{4mm} $\wt b\in \R^{m}$
\State \hspace{4mm} $L\in \R^{p\times d}$, $\lambda\geq 0$
\State {\bf end members}
\State
\Procedure{Initialize}{$A_1\in \R^{n_1\times d_1},\ldots,A_q\in \R^{n_q\times d_q}, \bar s\geq 1,C_{\mathrm{base}},T_{\mathrm{base}}, L\in \R^{p\times d}, \lambda\geq 0, b\in \R^n$}
\State Choose the sketching dimension $m$ from $\bar s$ as in Theorem~\ref{thm:app:spline}
\State $\textsc{DTT.Initialize}(A_1,\ldots,A_q, m,C_{\mathrm{base}},T_{\mathrm{base}})$
\State $A_1\gets A_1,\ldots,A_q\gets A_q$
\State $L\gets L$, $\lambda\gets \lambda$
\State Let $\Pi_q \in \R^{m\times n}$ be the sketching matrix corresponds to {\sc DTT}
\State $\wt b\gets \Pi_q b$
\EndProcedure
\State
\Procedure{Update}{$i\in [q], B\in \R^{n_i\times d_i}$}
\State $\textsc{DTT.Update}(i, B)$
\State $A_i\gets A_i+B$
\EndProcedure
\State
\Procedure{Query}{$ $}
\State $M\gets \textsc{DTT.}J_{0,\log q}$
\State $\wt x \gets \arg \min_{x \in \R^d} \|Mx - \wt b\|_2^2 + \lambda \|Lx\|_2^2$
\State \Return $\wt x$
\EndProcedure
\end{algorithmic}
\end{algorithm}

We are now ready to prove the main theorem.

\begin{theorem}[Formal version of Theorem~\ref{thm:spline}]\label{thm:app:spline}
Let $\epsilon\in (0,1/8]$ and $\delta\in (0,1)$. There exists an algorithm (Algorithm~\ref{alg:kronecker_spline}) with the following procedures:
\begin{itemize}
    \item \textsc{Initialize}$(A_1\in \R^{n_1\times d_1},\ldots,A_q\in \R^{n_q\times d_q}, \bar s\geq 1)$: Given matrices $A_1\in \R^{n_1\times d_1},\ldots,A_q\in \R^{n_q\times d_q}$ and an upper bound $\bar s\geq 1$ on the statistical dimension at the query times, the data structure chooses the sketching dimension $m$ from $\bar s$ as below and pre-processes in time $\wt O(s_C\sum_{i=1}^q \nnz(A_i) + qmd)$ plus the time to compute $\Pi^q b$, which is $O(q\cdot\nnz(b))$ for {\sc CountSketch} and {\sc TensorSketch} and $\wt O(qm\cdot\nnz(b))$ for {\sc OSNAP} and {\sc TensorSRHT}, where $s_C$ is the leaf sparsity of Theorem~\ref{thm:app:tree_init}. The choice $\bar s=d$ is always valid, since $\sd_{\lambda}(A,L)\leq d$ (each of the $p$ terms $1/(1+\lambda/\gamma_i^2)$ is at most $1$).
    \item \textsc{Update}$(i\in [q], B\in \R^{n_i\times d_i})$: Given an index $i\in [q]$ and an update matrix $B\in \R^{n\times d}$, the data structure updates in time $\wt O(s_C\nnz(B) + md)$.
    \item \textsc{Query}: Let $A^{(t)}$ denote the Tensor product $\bigotimes_{i=1}^q A_i$ of the current factors at time $t$. For every time $t$ at which $\sd_{\lambda}(A^{(t)}, L)\leq \bar s$, $\rank(L)=p$ and $\rank\left(\mat{c} A^{(t)} \\ L \rix\right)=d$, Query at time $t$ outputs a solution $\wt x \in \R^d$ with probability at least $1-2\delta$ such the $\wt x$ is an $\epsilon$ approximate to the Tensor Spline Regression problem i.e. $\|A^{(t)} \wt x - b\|_2^2 + \lambda \cdot \|L \wt x\|_2^2 \leq (1+ \epsilon) \cdot \opt$ where $\opt = \min_{x \in \R^d} \|A^{(t)}x - b\|_2^2 + \lambda \cdot \|L x\|_2^2$. Query takes time $\wt O(md^{(\omega - 1)} + pd^{(\omega - 1)} + d^{\omega})$.
\end{itemize}
Moreover, we have that if
\begin{itemize}
    \item $C_{\mathrm{base}}$ is {\sc CountSketch}, $T_{\mathrm{base}}$ is {\sc TensorSketch},  then $m = \Omega(\epsilon^{-1}q\bar s^2/\delta)$ and $s_C=1$.
    \item $C_{\mathrm{base}}$ is {\sc OSNAP}, $T_{\mathrm{base}}$ is {\sc TensorSRHT}, then $m=\wt \Omega(\epsilon^{-1}q \bar s^2\log^3(1/(\epsilon\delta)))$ and $s_C=\Theta(\sqrt{q}\cdot\bar s\log(1/\delta)/\sqrt{\epsilon})$.
\end{itemize}
\end{theorem}

\begin{proof}
{\bf Proof of Correctness for \textsc{Query}}.

Fix a time $t$ at which $\sd_{\lambda}(A^{(t)}, L)\leq \bar s$ and the two rank conditions hold, and write $A=A^{(t)}$. Let $S$ denote the sketching matrix related to the dynamic tree. To prove that it is enough to use sketching whose dimension depends on $\bar s$, we just need to show that $S$ satisfies the two guarantees of Lemma~\ref{lem:dssw18_1}. We will use approximate matrix product (Def.~\ref{def:amp}) to prove this guarantee. Let
\begin{align*}
    \eta = \sqrt{\epsilon/2}/\bar s,
\end{align*}
and consider an $(\eta,\delta)$-approximate matrix product sketching. By Lemma~\ref{lem:dssw18_2}, we have that $\|U_1\|_F^2=\sd_{\lambda}(A, L)\leq \bar s$. For the first part, we set $C=D=U_1$, then with probability at least $1-\delta$,
\begin{align*}
    \|U_1^\top S^\top S U_1-U_1^\top U_1\|_2 \leq & ~ \|U_1^\top S^\top S U_1-U_1^\top U_1\|_F \\
    \leq & ~ \eta \|U_1\|_F^2 \\
    = & ~ \sqrt{\epsilon/2}\cdot \sd_{\lambda}(A, L)/\bar s \\
    \leq & ~ \sqrt{\epsilon/2} \\
    \leq & ~ 1/4,
\end{align*}
where the fourth step uses $\sd_{\lambda}(A, L)\leq \bar s$, and the last step uses $\epsilon\leq 1/8$, so that $\epsilon/2\leq 1/16$. For the second part, we set $C=U_1$ and $D=b-Ax^*$. Since $\bar s\geq 1$ and $\sd_{\lambda}(A, L)\leq \bar s$, we have $\sqrt{\sd_{\lambda}(A, L)}\leq \max\{1,\sd_{\lambda}(A, L)\}\leq \bar s$. Moreover, $\|b-Ax^*\|_2^2\leq \|b-Ax^*\|_2^2+\lambda\|Lx^*\|_2^2={\rm OPT}$. Hence, with probability at least $1-\delta$,
\begin{align*}
    \|U_1^\top (S^\top S-I)(b-Ax^*) \|_2 \leq & ~ \|U_1^\top (S^\top S-I)(b-Ax^*) \|_F \\
    \leq & ~ \eta \|U_1\|_F \|b-Ax^*\|_2 \\
    = & ~ \sqrt{\epsilon/2}\cdot (\sqrt{\sd_{\lambda}(A,L)}/\bar s)\cdot \|b-Ax^*\|_2 \\
    \leq & ~ \sqrt{\epsilon/2}\cdot \sqrt{{\rm OPT}} \\
    = & ~ \sqrt{\epsilon~{\rm OPT}/2}.
\end{align*}
By a union bound over the two events, $S$ satisfies both guarantees of Lemma~\ref{lem:dssw18_1} with probability at least $1-2\delta$. Applying Lemma~\ref{lem:dssw18_1} with $2\delta$ in place of $\delta$, the output $\wt x$ of Query satisfies $\|A\wt x-b\|_2^2 + \lambda \|L\wt x\|_2^2 \leq (1+\epsilon){\rm OPT}$ with probability at least $1-2\delta$.

This means that to obtain good approximation for Spline regression, it is enough to employ $(\eta, \delta)$-approximate matrix product with $\eta=\sqrt{\epsilon/2}/\bar s$. Since $\eta^{-2}=2\bar s^2/\epsilon$, the dimension follows from Lemma~\ref{lem:cs_tensorsketch} and~\ref{lem:osnap_tensorsrht} with $\eta$ as the accuracy of the approximate matrix product, which gives $m=\Omega(\epsilon^{-1}q\bar s^2/\delta)$ for {\sc CountSketch} and {\sc TensorSketch}, and $m=\wt\Omega(\epsilon^{-1}q\bar s^2\log^3(1/(\epsilon\delta)))$ for {\sc OSNAP} and {\sc TensorSRHT} (since $\eta^{-1}\leq\sqrt{2}\,\bar s/\epsilon$, using accuracy $\eta$ inside the logarithm adds at most a factor $\sqrt{2}\,\bar s$ there, which is absorbed by $\wt\Omega$). In the second case, part 2 of Lemma~\ref{lem:osnap_tensorsrht} also needs leaf sparsity $s_C=\Omega(\sqrt{q}\log(1/\delta)/\eta)$, and since $\eta^{-1}=\sqrt{2}\,\bar s/\sqrt{\epsilon}$, the choice $s_C=\Theta(\sqrt{q}\cdot\bar s\log(1/\delta)/\sqrt{\epsilon})$ suffices.

We note that even though the definition of statistical dimension in~\cite{akk+20} and ours is different, our definition is more general and the~\cite{akk+20} definition can be viewed as picking $L^\top L$ to be the identity matrix. Hence, one can easily generalize the conclusion of Lemma~\ref{lem:cs_tensorsketch},~\ref{lem:osnap_tensorsrht} and~\ref{lem:srht_tensorsrht} to our definition of statistical dimension via the argument used in~\cite{anw17,dssw18,akk+20}.

{\bf Proof of \textsc{Initialization}}. The initialization part of Algorithm~\ref{alg:kronecker_spline} is identical to the initialization part of Algorithm~\ref{alg:kronecker_reg}. Therefore, by Theorem~\ref{thm:app:dynamic_k_reg}, initialization takes time $\wt O(s_C\sum_{i=1}^q \nnz(A_i) + qmd)$ plus the time to compute $\Pi^q b$ by one leaf-to-root pass per nonzero entry of $b$, which is $O(q\cdot\nnz(b))$ for {\sc CountSketch} and {\sc TensorSketch} and $\wt O(qm\cdot\nnz(b))$ for {\sc OSNAP} and {\sc TensorSRHT}.

{\bf Proof of \textsc{Update}}. We just update our DTT data structure. This takes time $\wt O(s_C\nnz(B) + md)$ from Theorem~\ref{thm:app:tree_init}.

{\bf Proof of \textsc{Query} time}. As mentioned in Section~\ref{sec:spline_regression}, we can compute $\wt x \gets \arg \min_{x \in \R^d} \|Mx - \wt b\|_2^2 + \lambda \|Lx\|_2^2$ by $\wt x \gets (M^\top M + \lambda L^\top L)^{-1} M^{\top} \wt b$. We can view Query as consisting of the following $5$ steps:

\begin{itemize}
    \item Initializing $M$ takes time $\wt O(md)$ since $M \in \R^{m \times d}$.
    \item $M^\top (\wt b)$, this takes $O(md)$ time because we are multiplying a $d \times m$ matrix with a $m \times 1$ vector.
    \item $M^\top \cdot M$, this step takes $\Tmat(d,m,d)$ time as we are multiplying a $d \times m$ matrix with an $m \times d$ matrix.
    \item $L^\top L$, this step takes $\Tmat(d,p,d)$ time as we are multiplying a $d \times p$ matrix with a $p \times d$ matrix.
    \item $M^\top M + \lambda L^\top L$, this takes $O(d^2)$ time as we are adding two $d \times d$ size matrices.
    \item $(M^\top M + \lambda L^\top L)^{-1}$, this step takes $\Tmat(d,d,d)$ time as it is inverting a matrix of size $d \times d$.
    \item $(M^\top M + \lambda L^\top L)^{-1} (M^\top \wt b)$. In this step, we are multiplying $(M^\top M + \lambda L^\top L)^{-1}$ which is a $d \times d$ matrix with a vector $M^\top \wt b$ of size $d$. This takes time $O(d^2)$.
\end{itemize}
Overall, the total running time of $\textsc{Query}$ is
\begin{align*}
    & ~ \wt O(md + md + \Tmat(d,m,d) + \Tmat(d,p,d) + d^2 + \Tmat(d,d,d) + d^2) \\
    = & ~ \wt O(md + md^{(\omega - 1)} + pd^{(\omega - 1)} + d^{\omega} + d^2) \\
    = & ~ \wt O(md^{(\omega - 1)} + pd^{(\omega - 1)} + d^{\omega})
\end{align*}
\end{proof}
\section{Omitted details and proofs for Dynamic Tensor Low Rank Approximation (section~\ref{sec:low_rank_approximation})}\label{sec:appendix-low-rank}

For low rank approximation, we show it is enough to design sketching matrix whose dimension depends on $k$ instead of $d$. 

\begin{definition}[Projection Cost Preserving Sketch]
Let $A\in \R^{n\times d}$, we say a sketching matrix $S\in \R^{m\times n}$ is a $k$ \emph{Projection Cost Preserving Sketch} (PCPSketch) with error $\epsilon$ if for any orthogonal projection $P\in \R^{d\times d}$ of rank $k$, we have
\begin{align*}
    (1-\epsilon) \|A-AP\|_F^2 \leq \|SA-SAP\|_F^2+c\leq (1+\epsilon) \|A-AP\|_F^2
\end{align*}
where $c$ is a fixed constant, possibly negative, independent of $P$, which may depend on $A$ and $SA$.
\end{definition}

For low rank approximation, we note that the optimal low rank approximation for $A$ is projecting onto the top $k$ right singular vectors. Hence, if we can prove that the sketching corresponds to the dynamic tree is a $k$-PCPSketch, then we are done. We will require a few technical tools.

\begin{lemma}[Lemma 6 and the proof of Theorem 12 (Section 7.1) of~\cite{cem+15}]
\label{lem:cem_1}
Let $A\in \R^{n\times d}$, and let $A_k$ be a best rank-$k$ approximation of $A$. Suppose $S\in \R^{m\times n}$ is drawn from a distribution that is an $(\epsilon, \delta, k, d, n)$-OSE, applied to the rank-$k$ matrix $A_k$ (that is, $S$ is an $\epsilon$-subspace embedding for the column space of $A_k$ with probability at least $1-\delta$), and has the $(\epsilon/\sqrt{k},\delta)$-approximate matrix product property. Then, with probability at least $1-O(\delta)$, $S$ is a $k$-PCPSketch for $A$ with error $O(\epsilon)$, where the constant $c$ may be negative.
\end{lemma}

\begin{proof}
Let $A_{\setminus k}=A-A_k$, let $r=\rank(A_k)\leq k$, and let $W\in \R^{n\times r}$ have orthonormal columns spanning the column space of $A_k$. Since $A_k$ is a best rank-$k$ approximation, it is the projection of $A$ onto its own column space, so $A_k=WW^\top A$. Hence $W^\top A_{\setminus k}=W^\top A-W^\top WW^\top A=0$, and so $A_k^\top A_{\setminus k}=A^\top W W^\top A_{\setminus k}=0$. Let $E_1=A_k^\top S^\top SA_k-A_k^\top A_k$, $E_2=A_{\setminus k}^\top S^\top SA_{\setminus k}-A_{\setminus k}^\top A_{\setminus k}$ and $E_3=A_k^\top S^\top SA_{\setminus k}$. Since $A^\top A=A_k^\top A_k+A_{\setminus k}^\top A_{\setminus k}$, we have $A^\top S^\top SA=A^\top A+E_1+E_2+E_3+E_3^\top$. By a union bound, with probability at least $1-3\delta$, $S$ is an $\epsilon$-subspace embedding for the column space of $A_k$, so that $-\epsilon A_k^\top A_k\preceq E_1\preceq \epsilon A_k^\top A_k$, and the approximate matrix product guarantee holds for the pairs $(A_{\setminus k},A_{\setminus k})$ and $(W,A_{\setminus k})$. For the second pair we pad $W$ with $d-r$ zero columns, so that both matrices have $d$ columns as in Definition~\ref{def:amp}. This changes neither $\|W\|_F$ nor the nonzero rows of the products. We work on this event.

Fix an orthogonal projection $P\in \R^{d\times d}$ of rank $k$, let $Y=I-P$, and set $c=-\tr(E_2)$, which does not depend on $P$. We have $\|SA-SAP\|_F^2=\tr(YA^\top S^\top SAY)$ and $\|A-AP\|_F^2=\tr(YA^\top AY)$. Using $Y^2=Y$ and $P^2=P$, we get $\tr(YE_2Y)-\tr(E_2)=\tr(E_2Y)-\tr(E_2)=-\tr(E_2P)=-\tr(PE_2P)$, and $\tr(YE_3^\top Y)=\tr(YE_3Y)$ because the two matrices are transposes of each other. Therefore
\begin{align*}
    \|SA-SAP\|_F^2+c = \|A-AP\|_F^2+\tr(YE_1Y)-\tr(PE_2P)+2\tr(YE_3Y).
\end{align*}
We bound each error term by $\epsilon\|A-AP\|_F^2$. We use $\|A-AP\|_F^2=\|A_kY\|_F^2+\|A_{\setminus k}Y\|_F^2$, which holds because the cross term $2\tr(YA_k^\top A_{\setminus k}Y)$ vanishes, and $\|A-AP\|_F\geq \|A_{\setminus k}\|_F$, which holds because $AP$ has rank at most $k$.
\begin{itemize}
    \item From $-\epsilon A_k^\top A_k\preceq E_1\preceq \epsilon A_k^\top A_k$, we get $-\epsilon YA_k^\top A_kY\preceq YE_1Y\preceq \epsilon YA_k^\top A_kY$, so $|\tr(YE_1Y)|\leq \epsilon\tr(YA_k^\top A_kY)=\epsilon\|A_kY\|_F^2\leq \epsilon\|A-AP\|_F^2$.
    \item Write $P=ZZ^\top$ with $Z\in \R^{d\times k}$ having orthonormal columns. Then $\tr(PE_2P)=\tr(Z^\top E_2Z)$, and by the Cauchy--Schwarz inequality $|\tr(Z^\top E_2Z)|\leq \|Z\|_F\|E_2Z\|_F\leq \sqrt{k}\|E_2\|_F$, since $\|Z\|_F=\sqrt{k}$ and $\|E_2Z\|_F\leq \|E_2\|_F\|Z\|_2=\|E_2\|_F$. The approximate matrix product guarantee for $(A_{\setminus k},A_{\setminus k})$ gives $\|E_2\|_F\leq (\epsilon/\sqrt{k})\|A_{\setminus k}\|_F^2$, so $|\tr(PE_2P)|\leq \epsilon\|A_{\setminus k}\|_F^2\leq \epsilon\|A-AP\|_F^2$.
    \item Since $A_k^\top=A^\top WW^\top$, we have $E_3=A^\top WG$ with $G=W^\top S^\top SA_{\setminus k}=W^\top S^\top SA_{\setminus k}-W^\top A_{\setminus k}$. The approximate matrix product guarantee for $(W,A_{\setminus k})$ gives $\|G\|_F\leq (\epsilon/\sqrt{k})\|W\|_F\|A_{\setminus k}\|_F=\epsilon\sqrt{r/k}\,\|A_{\setminus k}\|_F\leq \epsilon\|A_{\setminus k}\|_F$. Hence $|\tr(YE_3Y)|=|\tr((W^\top AY)^\top GY)|\leq \|W^\top AY\|_F\|GY\|_F\leq \|AY\|_F\cdot \epsilon\|A_{\setminus k}\|_F\leq \epsilon\|A-AP\|_F^2$, where we used $\|W^\top\|_2=1$, $\|Y\|_2\leq 1$ and $\|AY\|_F=\|A-AP\|_F$.
\end{itemize}
Summing the three bounds, with the last one counted twice, $\big|\|SA-SAP\|_F^2+c-\|A-AP\|_F^2\big|\leq 4\epsilon\|A-AP\|_F^2$ for every orthogonal projection $P$ of rank $k$. So $S$ is a $k$-PCPSketch for $A$ with error $4\epsilon$ with probability at least $1-3\delta$. Because $c$ may be negative, the argument needs no condition on $\tr(E_2)$, unlike Lemma 6 of~\cite{cem+15}.
\end{proof}

\begin{algorithm}[!ht]\caption{Our low rank approximation algorithm}
\label{alg:kronecker_low_rank}
\begin{algorithmic}[1]
\State {\bf data structure} {\sc LowRankMaintenance}
\State
\State {\bf members}
\State \hspace{4mm} $\textsc{DynamicTensorTree DTT} $
\State \hspace{4mm} $A_1\in \R^{n_1\times d_1},\ldots,A_q\in \R^{n_q\times d_q}$
\State {\bf end members}
\State 
\Procedure{Initialize}{$A_1\in \R^{n_1\times d_1},\ldots,A_q\in \R^{n_q\times d_q}$}
\State $\textsc{DTT.Initialize}(A_1,\ldots,A_q)$
\State $A_1\gets A_1,\ldots,A_q\gets A_q$
\EndProcedure
\State
\Procedure{Update}{$i\in [q], B\in \R^{n_i\times d_i}$}
\State $\textsc{DTT.Update}(i, B)$
\State $A_i\gets A_i+B$
\EndProcedure
\State
\Procedure{Query}{}
\State $M\gets \textsc{DTT.}J_{0,\log q}$
\State Compute SVD of $M$ such that $M=U\Sigma V^\top$
\State $U_k\gets \text{top $k$ right singular vectors of $M$}$
\State \Return $A_1\otimes \ldots \otimes A_q U_k^\top U_k$ in factored form
\EndProcedure
\end{algorithmic}
\end{algorithm}

\begin{theorem}[Formal version of Theorem~\ref{thm:low_rank}]\label{thm:app:low_rank}
Let $\epsilon\in (0,1]$ and $\delta\in (0,1)$. There exists an algorithm (Algorithm~\ref{alg:kronecker_low_rank}) that has the following procedures
\begin{itemize}
    \item \textsc{Initialize}$(A_1\in \R^{n_1\times d_1},\ldots,A_q\in \R^{n_q\times d_q})$: Given matrices $A_1\in \R^{n\times d},\ldots,A_q\in \R^{n\times d}$, the data structure processes in time $\wt O(s_C\sum_{i=1}^q \nnz(A_i) + qmd)$, where $s_C$ is the leaf sparsity of Theorem~\ref{thm:app:tree_init}.
    \item \textsc{Update}$(i\in [q], B\in \R^{n_i\times d_i})$: Given an index $i\in [q]$ and an update matrix $B\in \R^{n\times d}$, the data structure updates the approximation for $A_1\otimes \ldots \otimes (A_i+B)\otimes \ldots \otimes A_q$ in time $\wt O(s_C\nnz(B) + md)$.
    \item \textsc{Query}: Let $A$ denote the tensor product $\bigotimes_{i=1}^q A_i$. The data structure outputs a rank-$k$ approximation $C$ such that, with probability at least $1-O(\delta)$,
    \begin{align*}
        \|C-A\|_F \leq & ~ (1+\epsilon)\min_{\mathrm{rank}-k~A'}~\|A'-A\|_F.
    \end{align*}
    The time to output $C$ is $\wt O(md^{\omega-1})$.
\end{itemize}
Moreover, we have that if
\begin{itemize}
    \item $C_{\mathrm{base}}$ is {\sc CountSketch}, $T_{\mathrm{base}}$ is {\sc TensorSketch}, $m=\Omega(\epsilon^{-2}qk^2 1/\delta)$ and $s_C=1$. 
    \item $C_{\mathrm{base}}$ is {\sc OSNAP}, $T_{\mathrm{base}}$ is {\sc TensorSRHT}, $m=\wt \Omega(\epsilon^{-2}qk^2 \log^3(1/(\epsilon\delta)))$ and $s_C=\Theta(\sqrt{q}\cdot k\log(1/\delta)/\epsilon)$.
\end{itemize}
\end{theorem}

\begin{proof}
{\bf Proof of Correctness for \textsc{Query}}.

Let $S$ denote the sketching matrix related to the dynamic tree, so that $M=SA$ in Query. By Lemma~\ref{lem:cem_1} applied with $\epsilon$ replaced by $\epsilon/c_0$ for an absolute constant $c_0\geq 3$, chosen so that the error $O(\epsilon/c_0)$ of the lemma is at most $\epsilon/3$ (this changes $m$ only by a constant factor), $S$ is a $k$-PCPSketch for $A$ with error $\epsilon/3$ with probability at least $1-O(\delta)$, provided it is an $(\epsilon/c_0,\delta,k,d,n)$-OSE and has the $(\epsilon/(c_0\sqrt{k}),\delta)$-approximate matrix product property. Choosing the dimension according to Lemma~\ref{lem:cs_tensorsketch} and~\ref{lem:osnap_tensorsrht} gives both properties for the values of $m$ in the theorem. For {\sc OSNAP} and {\sc TensorSRHT}, part 1 of Lemma~\ref{lem:osnap_tensorsrht} with $\sd_\lambda=k$ and accuracy $\epsilon/c_0$ needs $m=\wt\Omega(\epsilon^{-2}qk^2\log^3(1/(\epsilon\delta)))$ and $s_C=\Omega(\sqrt{q}\cdot k\log(1/\delta)/\epsilon)$, and part 2 with accuracy $\epsilon/(c_0\sqrt{k})$ needs $m=\wt\Omega(\epsilon^{-2}qk\log^3(1/(\epsilon\delta)))$ (the factor $c_0\sqrt{k}$ inside the logarithm is absorbed by $\wt\Omega$) and $s_C=\Omega(\sqrt{q}\cdot \sqrt{k}\log(1/\delta)/\epsilon)$. Since $k\geq \sqrt{k}$ and $k^2\geq k$, the conditions of part 1 imply those of part 2, so the choice of $m$ and $s_C$ in the theorem gives both properties.

We now show that a $k$-PCPSketch for $A$ with error $\epsilon/3$ gives a matrix $C$ that is a good rank-$k$ approximation of $A$. Recall that for any orthogonal projection $P\in \R^{d\times d}$ of rank $k$, we have
\begin{align*}
    (1-\epsilon/3) \|A-AP\|_F^2 \leq \|SA-SAP\|_F^2+c \leq (1+\epsilon/3) \|A-AP\|_F^2,
\end{align*}
where the constant $c$ does not depend on $P$ and may be negative. We need to design the projection $P$. Let $SA=U\Sigma V^\top$ be the singular value decomposition of $SA$, and let $U_k\in \R^{k\times d}$ be the matrix whose rows are the top $k$ right singular vectors of $SA$. Set $P=U_k^\top U_k$, we note that $SAP=SAU_k^\top U_k$ is the optimal rank-$k$ approximation for $SA$, so $\|SA-SAP\|_F^2\leq \|SA-SAP'\|_F^2$ for every orthogonal projection $P'$ of rank $k$. On the other hand, let $Q\in \R^{d\times d}$ be an orthogonal projection of rank $k$ such that $AQ$ is the optimal rank-$k$ approximation for $A$, so that $\|A-AQ\|_F=\mathrm{OPT}_k\coloneqq\min_{\mathrm{rank}\text{-}k~A'}\|A'-A\|_F$. We have
\begin{align*}
    (1-\epsilon/3)\|A-AP\|_F^2 \leq & ~ \|SA-SAP\|_F^2+c \\
    \leq & ~ \|SA-SAQ\|_F^2+c \\
    \leq & ~ (1+\epsilon/3)\|A-AQ\|_F^2 \\
    = & ~ (1+\epsilon/3)\mathrm{OPT}_k^2,
\end{align*}
where the first and third steps are the PCPSketch guarantee for $P$ and for $Q$, and the second step holds because $P$ is optimal for $SA$. The same $c$ appears on both sides of the second step, so the sign of $c$ is never used. Since $(1+\epsilon)(1-\epsilon/3)=1+2\epsilon/3-\epsilon^2/3\geq 1+\epsilon/3$ for $\epsilon\in(0,1]$, we have $(1+\epsilon/3)/(1-\epsilon/3)\leq 1+\epsilon$. Hence the output $C=AP=AU_k^\top U_k$ satisfies
\begin{align*}
    \|C-A\|_F \leq \sqrt{\frac{1+\epsilon/3}{1-\epsilon/3}}\cdot \mathrm{OPT}_k \leq \sqrt{1+\epsilon}\cdot\mathrm{OPT}_k \leq (1+\epsilon)\mathrm{OPT}_k
\end{align*}
with probability at least $1-O(\delta)$.

{\bf Proof of \textsc{Initialization}}. The initialization part of Algorithm~\ref{alg:kronecker_low_rank} runs \textsc{DTT.Initialize} and stores the matrices $A_1,\ldots,A_q$. By Theorem~\ref{thm:app:tree_init}, the first step takes time $\wt O(s_C\sum_{i=1}^q \nnz(A_i) + qmd)$, and the second step takes time $O(\sum_{i=1}^q \nnz(A_i))$. Since $s_C\geq 1$, the time for initialization is $\wt O(s_C\sum_{i=1}^q \nnz(A_i) + qmd)$.

{\bf Proof of \textsc{Update}}. We just update our DTT data structure. This takes time $\wt O(s_C\nnz(B) + md)$ from Theorem~\ref{thm:app:tree_init}.

{\bf Proof of \textsc{Query} time}. We can view Query as consisting of the following $5$ steps:

\begin{itemize}
    \item Initializing $M$ takes time $\wt O(md)$ since $M \in \R^{m \times d}$.
    \item Computing the SVD of $M$ takes time $O(md^{\omega-1})$ since $M$ is $m \times d$ matrix and $m\geq d$.
    \item $U_k$ can be obtained directly from the SVD computation so this step doesn't take any additional time.
    \item Once the algorithm has $U_k$, it can terminate as it outputs $(A_1,...,A_q,U_k)$ in factored form.
\end{itemize}
Overall, the total running time of $\textsc{Query}$ is
\begin{align*}
    \wt O(md + md^{\omega-1}))
    = & ~ \wt O(md^{\omega-1}). \qedhere
\end{align*}

\end{proof}

\end{document}